%% file: seigen-tiling.tex
\documentclass[format=acmsmall, review=false, screen=true]{acmart}

\usepackage{booktabs} 

\usepackage{subcaption} 

\usepackage{listings}  
\usepackage[ruled,linesnumbered]{algorithm2e} 

\SetAlFnt{\small}
\SetAlCapFnt{\small}
\SetAlCapNameFnt{\small}
\SetAlCapHSkip{0pt}
\IncMargin{-\parindent}
\SetKwInput{kwInput}{Input}
\SetKwInput{kwOutput}{Output}
\SetKw{Not}{not}
\SetKw{In}{in}
\SetKw{True}{true}
\SetKw{False}{false}
\SetKw{Skip}{skip}
\SetKwRepeat{Do}{do}{while}
\SetKwComment{Comment}{// }{}
\let\oldnl\nl
\newcommand{\nonl}{\renewcommand{\nl}{\let\nl\oldnl}}

\usepackage{tabulary} 

\usepackage{multicol}
\usepackage{multirow}
\usepackage{boldline}
\usepackage{pifont}

\usepackage{array}
\newcolumntype{P}[1]{>{\centering\arraybackslash}p{#1}}
\newcolumntype{M}[1]{>{\centering\arraybackslash}m{#1}}
\newcolumntype{N}{@{}m{0pt}@{}}

\setcopyright{acmlicensed}

\usepackage{fancybox}
\makeatletter
\newenvironment{CenteredBox}{%
\begin{Sbox}}{
\end{Sbox}\centerline{\parbox{\wd\@Sbox}{\TheSbox}}}
\makeatother

\begin{document}

\setlength{\abovedisplayskip}{3pt}
\setlength{\belowdisplayskip}{3pt}
\setlength{\abovedisplayshortskip}{0pt}
\setlength{\belowdisplayshortskip}{0pt}

\title[Automated Tiling of Unstructured Mesh Computations]{Automated Tiling of Unstructured Mesh Computations with Application to Seismological Modelling}
\author{Fabio Luporini}
\affiliation{%
  \institution{Imperial College London}
}
\author{Michael Lange}\affiliation{\institution{Imperial College London}}
\author{Christian T. Jacobs}\affiliation{\institution{University of Southampton}}
\author{Gerard J. Gorman}\affiliation{\institution{Imperial College London}}
\author{J. Ramanujam}\affiliation{\institution{Louisiana State University}}
\author{Paul H. J. Kelly}\affiliation{\institution{Imperial College London}}

\begin{abstract}
Sparse tiling is a technique to fuse loops that access common data, thus increasing data locality. Unlike traditional loop fusion or blocking, the loops may have different iteration spaces and access shared datasets through indirect memory accesses, such as {\tt A[map[i]]} -- hence the name ``sparse''. One notable example of such loops arises in discontinuous-Galerkin finite element methods, because of the computation of numerical integrals over different domains (e.g., cells, facets). The major challenge with sparse tiling is implementation -- not only is it cumbersome to understand and synthesize, but it is also onerous to maintain and generalize, as it requires a complete rewrite of the bulk of the numerical computation. In this article, we propose an approach to extend the applicability of sparse tiling based on raising the level of abstraction. Through a sequence of compiler passes, the mathematical specification of a problem is progressively lowered, and eventually sparse-tiled C for-loops are generated. Besides automation, we advance the state-of-the-art by introducing: a revisited, more efficient sparse tiling algorithm; support for distributed-memory parallelism; a range of fine-grained optimizations for increased run-time performance; implementation in a publicly-available library, SLOPE; and an in-depth study of the performance impact in Seigen, a real-world elastic wave equation solver for seismological problems, which shows speed-ups up to 1.28$\times$ on a platform consisting of 896 Intel Broadwell cores.
\end{abstract}

%
%

%
%


\thanks{This work was supported by the Engineering and Physical Sciences Research Council through grants EP/I00677X/1, EP/L000407/1, EP/I012036/1], by the Imperial College London Department of Computing, and by the Imperial College London Intel Parallel Computing Centre (IPCC). The work of J. Ramanujam is supported by the US National Science Foundation award CCF-1619303, the Louisiana Board of Regents contract LEQSF(2016-19)-RD-B-03 and by Louisiana State University. The authors would like to thank the HPC Service Support team at Imperial College London for their help with the Helen cluster. The authors would also like to thank Gheorghe-Teodor Bercea, Lawrence Mitchell, and David Ham for their suggestions during the development of this project..}

\maketitle

\renewcommand{\shortauthors}{F. Luporini et al.}

\input{body}

\end{document}

%% file: body.tex
\section{Introduction}
In many unstructured mesh applications, for example those approximating the solution of partial differential equations (PDEs) using the finite volume or the finite element method, sequences of numerical operators accessing common fields need to be evaluated. Usually, these operators are implemented by iterating over sets of mesh elements and computing a kernel in each element. In languages such as C or Fortran, the resulting sequence of loops is typically characterized by heterogeneous iteration spaces and accesses to shared datasets (reads, writes, increments) through indirect pointers, like {\tt A[map[i]]}. One notable example of such operators/loops arises in discontinuous-Galerkin finite element methods, in which numerical integrals over different domains (e.g., cells, facets) are evaluated; here, {\tt A} could represent a discrete function, whereas {\tt map} could store connectivity information (e.g., from mesh elements to degrees of freedom). In this article, we devise compiler theory and technology to automate a sophisticated version of {\it sparse tiling}, a technique to maximize data locality when accessing shared fields (like the {\tt A} and {\tt map} arrays in the earlier example), which consists of fusing a sequence of loops by grouping iterations such that all data dependencies are honored. The goal is to improve the overall application performance with minimal disruption (none, if possible) to the source code. 


Three motivating real-world applications for this work are Hydra, Volna and Seigen. Hydra~\citep{hydra-op2} is a finite-volume computational fluid dynamics application used at Rolls Royce for the simulation of next-generation components of jet engines. Volna~\citep{ST-volna} is a finite-volume computational fluid dynamics application for the modelling of tsunami waves. Seigen~\citep{Seigen-paper} aims to solve the elastic wave equation using the discontinuous Galerkin finite element method for seismic exploration purposes. All these applications are characterized by the presence of a time-stepping loop, in which several loops over the mesh (thirty-three in Hydra, ten in Volna, twenty-five in Seigen) are repeatedly executed. These loops are characterized by the irregular dependence structure mentioned earlier, with for example indirect increments in one loop (e.g., {\tt A[m[i]] += f(...)}) followed by indirect reads in one of the subsequent loops (e.g., {\tt b = g(A[n[j]])}). The performance achievable by Seigen through sparse tiling will extensively be studied in Section~\ref{sec:performance}.

Although our work is general in nature, we are particularly interested in supporting increasingly sophisticated seismological problems that will be developed on top of Seigen. This has led to the following strategic decisions:
\begin{description}
\item[Automation, but no interest in legacy codes] Sparse tiling is an ``extreme optimization''. An implementation in a low level language requires a great deal of effort, as a thoughtful restructuring of the application is necessary. In common with many other low level transformations, it also makes the source code impenetrable, affecting maintenance and extensibility. We therefore aim for a fully automated system based on domain-specific languages (DSLs), which abstracts sparse tiling through a simple interface (i.e., a single construct to define a scope of fusible loops) and a tiny set of parameters for performance tuning (e.g., the tile size). We are not interested in automating sparse tiling in legacy codes, in which the key computational aspects (e.g., mesh iteration, distributed-memory parallelism) are usually hidden for software modularity, thus making such a transformation almost impossible. 
\item[Unstructured meshes require mixed static/dynamic analysis] Unstructured meshes are often used to discretize the computational domain, since they allow for an accurate representation of complex geometries. Their connectivity is stored by means of adjacency lists (or equivalent data structure), which leads to indirect memory accesses within the loop nests. Indirections break static analysis, thus making purely compiler-based approaches insufficient. Runtime data dependence analysis is essential for sparse tiling, so integration of compiler and run-time tracking algorithms becomes necessary.
\item[Realistic datasets not fitting in a single node] Real-world simulations often operate on terabytes of data, hence execution on multi-node systems is often required. We have extended the original sparse tiling algorithm to enable distributed-memory parallelism.
\end{description}

Sparse tiling does {\it not} change the semantics of a numerical method -- only the order in which some iterations are executed. Therefore, if most sections of a PDE solver suffer from computational boundedness and standard optimizations such as vectorization have already been applied, then sparse tiling, which targets memory-boundedness, will only provide marginal benefits (if any). Likewise, if a global reduction is present in between two loops, then there is no way for sparse tiling to be applied, unless the numerical method itself is rethought. This is regardless of whether the reduction is explicit (e.g., the first loop updates a global variable that is read by the second loop) or implicit (i.e., within an external function, as occurs for example in most implicit finite element solvers). These are probably the two greatest limitations of the technique; otherwise, sparse tiling may provide substantial performance benefits.

The rest of the article is structured as follows: in Section~\ref{sec:tiling:lc} we present the abstraction on which sparse tiling relies. We then show, in Section~\ref{sec:examples}, examples of how the algorithm works on shared- and distributed-memory systems. This is followed by the formalization of the algorithms (Sections~\ref{sec:data-dep-analysis},~\ref{sec:algorithm}) and the implementation of the compiler that automates sparse tiling (Section~\ref{sec:implementation}). The experimentation is described in Section~\ref{sec:performance}. A discussion on the limitations of the algorithms and the future work that we expect to carry out in the years to come conclude the article.

\section{The Loop Chain Abstraction for Unstructured Mesh Applications}
\label{sec:tiling:lc}

The {\em loop chain} is an abstraction introduced in~\cite{ST-KriegerHIPS2013}. Informally, a loop chain is a sequence of loops with no global synchronization points, enriched with information to enable run-time data dependence analysis -- necessary since indirect memory accesses inhibit common static approaches to loop optimization. The idea is to replace static with dynamic analysis, exploiting the information carried by a loop chain. Loop chains must somehow be added to or automatically derived (e.g., exploiting a DSL) from the input code. A loop chain will then be used by an {\em inspector/executor} scheme~\citep{ST-Saltz91}. The {\em inspector} is an algorithm performing data dependence analysis using the information carried by the loop chain, which eventually produces a {\em sparse tiling schedule}. This schedule is used by the {\em executor}, a piece of code semantically equivalent to the original sequence of loops (i.e., computing the same result) executing the various loop iterations in a different order.

Before diving into the description of the loop chain abstraction, it is worth observing two aspects.
\begin{itemize}
\item The inspection phase introduces an overhead. In many scientific computations, the data dependence pattern is static -- or, more informally, ``the topology does not change over time''. This means that the inspection cost may be amortized over multiple iterations of the executor. If instead the mesh changes over time (e.g., in case of adaptive mesh refinement), a new inspection must be performed. 
\item To adopt sparse tiling in a code there are two options. One possibility is to provide a library and leave the application specialists with the burden of carrying out the implementation (re-implementation in case of legacy code). A more promising alternative consists of raising the level of abstraction: programs can be written using a DSL; loop chain, inspector, and executor can then be automatically derived at the level of the intermediate representation. As we shall see in Section~\ref{sec:implementation}, the tools developed in this article enable both approaches, though our primary interest is in the automated approach (i.e., via DSLs).
\end{itemize}
These points will be further elaborated in later sections.

The loop chain abstraction was originally defined as follows:
\begin{itemize}
\item A loop chain $\mathbb{L} = [L_0, L_1, ..., L_{n-1}]$ is an ordered sequence of $n$ loops. There are no global synchronization points in between the loops. Although there may be dependencies between successive loops in the chain, the execution order of a loop's iterations does not influence the result.
\item $\mathbb{D} = \lbrace D_0, D_1, ..., D_{m-1} \rbrace$ is a collection of $m$ disjoint data spaces. Each loop accesses (reads from, writes to) a subset of these data spaces. An access can be either direct (e.g., {\tt A[i]}) or indirect (e.g., {\tt A[map(i)]}).
\item $R_{L_l\rightarrow D_d}(i)$ and $W_{L_l\rightarrow D_d}(i)$ are access relations for a loop $L_l$ over a data space $D_d \in \mathbb{D}$. They indicate which locations in the data space $D_d$ an iteration $i \in L_l$ reads from and writes to, respectively. A loop chain must provide all access relations for all loops.
\end{itemize}

\begin{figure}
\begin{CenteredBox}
\includegraphics[scale=0.6]{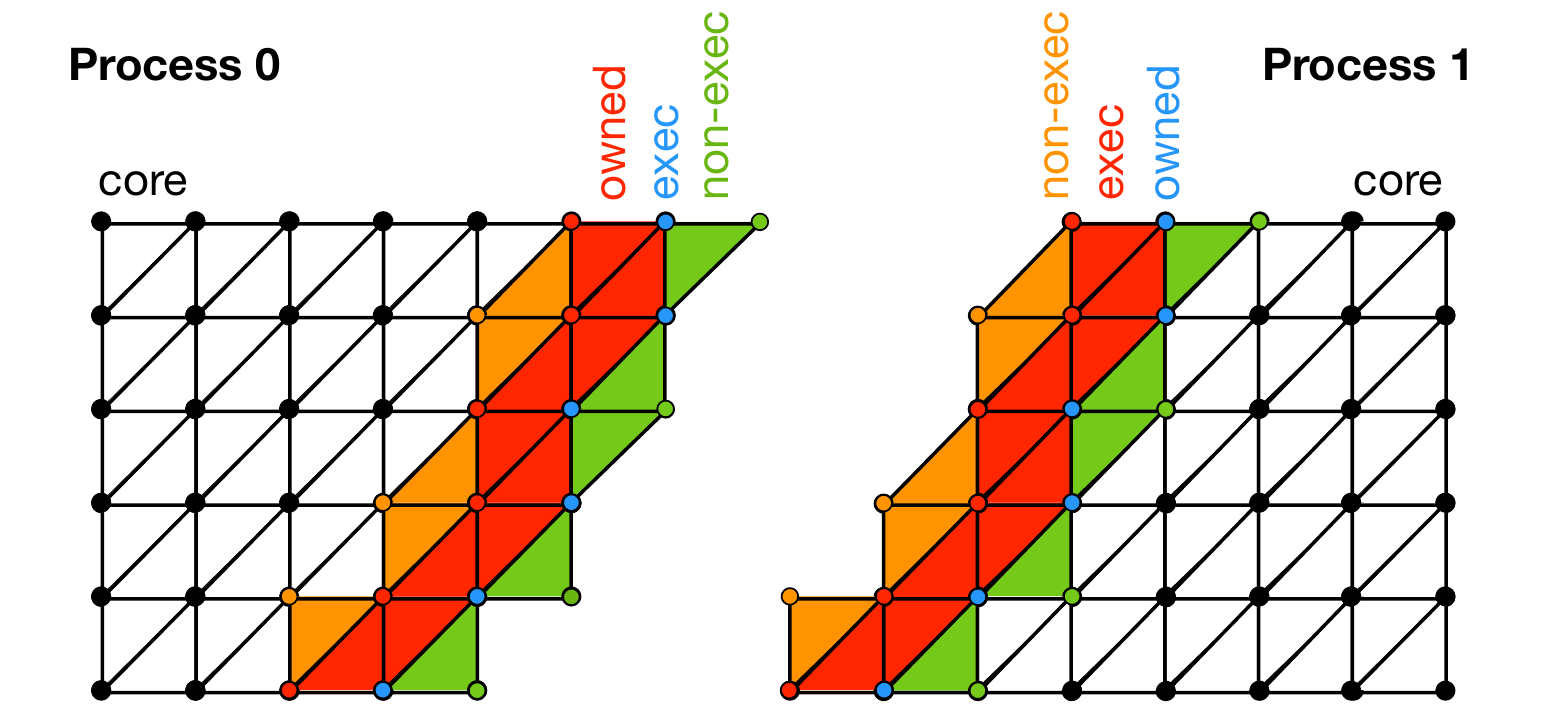}
\end{CenteredBox}
\caption{Two partitions of a mesh distributed to two neighboring processes, $P_0$ and $P_1$. The {\em core} region includes all iterations that can be processed without reading halo data. The {\em owned} iterations can be processed only by reading halo data. {\em Exec} is the set of iterations that must be executed because they indirectly write (increment) the owned iterations. The union of the {\em owned} and {\em exec} regions is referred to as the {\em boundary}. The {\em non-exec} region includes halo data which is indirectly read during the {\em exec} computation. The iterations in $P_0$'s ($P_1$'s) {\em exec} region are, logically, the same iterations in $P_1$'s ($P_0$'s) {\em owned} region; thus, we say that these iterations are ``redundantly executed''. Matching colors across the two processes represent identical subsets of iterations in the non-partitioned mesh. The image was inspired by an example in~\cite{florian-thesis}.}
\label{fig:sets}
\end{figure}

We here refine this definition, and specialize it for unstructured mesh applications. This allows the introduction of new concepts, necessary to extend the sparse tiling algorithm presented in~\cite{st-paper}. Some terminology and ideas are inspired by the programming model of OP2, a library for unstructured mesh applications~\citep{op2-main} used to implement the already mentioned Hydra code.

\begin{itemize}
\item A loop chain $\mathbb{L} = [L_0, L_1, ..., L_{n-1}]$ is an ordered sequence of $n$ loops. There are no global synchronization points in between the loops. Although there may be dependencies between successive loops in the chain, the execution order of a loop's iterations does not influence the result.

\item $\mathbb{S} = \lbrace S_0, S_1, ..., S_{m-1} \rbrace$ is a collection of $m$ disjoint iteration spaces. Possible iteration spaces are the topological entities of the mesh (e.g., cells, vertices) or the degrees of freedom associated with a function.

When using distributed-memory parallelism, an iteration space $S$ is logically split into three contiguous regions: {\em core}, {\em boundary}, and {\em non-exec} (see also Figure~\ref{fig:sets}). Given a generic process $P$ executing a loop over $S$, these regions represent:
\begin{description}
\item[core] the subset of iterations computed by $P$ that does not depend on halo exchanges. In other words, these are $P$'s local iterations.
\item[boundary] the union of two sub-regions, {\em owned} and {\em exec}, which are defined next. The {\em boundary} region requires up-to-date halo data. Like {\em core}, {\em owned} contains iterations owned by $P$; the data produced by {\em owned} are sent out through a halo exchange. The {\em exec} iterations, instead, are executed because they indirectly write (increment) data in $P$'s {\em owned} sub-region.
\item[non-exec] the subset of iterations not computed by $P$ mapping read-only data sent over to $P$ during a halo exchange.
\end{description}
 
An iteration space is uniquely identified by a name and the sizes of its three regions.

\item The {\em depth} is an integer indicating the extent of the boundary region. This is constant across all iteration spaces in $\mathbb{S}$.

\item $\mathbb{M} = \lbrace M_0, M_1, ..., M_{o-1} \rbrace$ is a set of $o$ maps. A map of arity $a$ is a vector-valued function $M : S_i \rightarrow S_j^a$ connecting elements in different iteration spaces. For example, we can express the mapping of a triangular cell $c$ to three vertices $v_0,v_1,v_2$ as $M(c) = [v_0,\ v_1,\ v_2]$; here cells and vertices are iteration spaces, while $c, v_0, v_1, v_2$ are iteration identifiers (i.e., natural numbers).

\item A loop $L_i$ over the iteration space $S$ is associated with one or more descriptors. A descriptor is a 2-tuple ${<}M,\ {\tt mode}{>}$. $M$ is either a map from $S$ to some other iteration spaces or the special placeholder $\perp$. In the former case, $L_i$ is accessing data associated with $M(S)$ indirectly; in the latter case, the data accesses are direct. ${\tt mode}$ is one of $[r,\ w,\ i]$, indicating whether a memory access is a read, write or increment.
\end{itemize}

There are a few crucial differences in this refined definition for the unstructured mesh case. One of them is the presence of iteration spaces in place of data spaces. In unstructured mesh applications, loops tend to access multiple data spaces associated with the same iteration space. A key observation is that if a loop is writing to some data spaces, then it is extremely likely that at least a subset of them will be accessed by the subsequent loop in the chain. The idea, therefore, is to rely on iteration spaces, rather than data spaces, to perform dependence analysis. This can substantially reduce the inspection cost, since typically $|\mathbb{S}| << |\mathbb{D}|$. Obviously, this relaxation might also create ``false dependences'', thus potentially affecting data communication. This would be the case if, for example, two consecutive, independent loops accessed different data fields associated with the same iteration space (e.g., {\it pressure} and {\it velocity} defined over the same set of degrees of freedom). In our experience, however, this rarely happens in practice (never in the case of the already mentioned Volna, Hydra and Seigen).

Another fundamental addition is the characterization of iteration spaces into the three regions core, boundary and non-exec. As we shall see, this separation is essential to enable distributed-memory parallelism. The extent of the boundary regions is captured by the {\em depth} of the loop chain. Informally, the {\em depth} tells how many extra ``strips'' of elements are provided by the neighboring processes. This allows some redundant computation along the partition boundary and also limits the depth of the loop chain (i.e., how many loops can be fused). The role of the parameter {\em depth} will be clear by the end of Section~\ref{sec:algorithm}.

\begin{figure}
\begin{CenteredBox}
\begin{subfigure}{0.46\textwidth}
\centering
\begin{lstlisting}[basicstyle=\tiny\ttfamily,keywordstyle=\ttfamily]
for t = 0 to T {

  // $L_0$: loop over edges, increment vertices
  for e = 0 to E {
    x = X + e;
    tmp_0 = edges2vertices[e + 0];
    tmp_1 = edges2vertices[e + 1];
    kernel1(x, tmp$\_$0, tmp$\_$1);
  }
  
  // $L_1$: loop over cells, increment vertices
  for c = 0 to C {
    res = R + c;
    tmp_0 = cells2vertices[c + 0];
    tmp_1 = cells2vertices[c + 1];
    tmp_2 = cells2vertices[c + 2];
    kernel2(res, tmp_0, tmp_1, tmp_2);
  }
  
  // $L_2$: loop over edges, read vertices
  for e = 0 to E {
    tmp_0 = edges2vertices[e + 0];
    tmp_1 = edges2vertices[e + 1];
    kernel3(tmp$\_$0, tmp$\_$1);
  }
}
\end{lstlisting}
\subcaption{Example sequence of sparse-tilable loops.}
\label{code:tiling-runningexample}
\end{subfigure}
\hspace{8mm}%
\begin{subfigure}{0.47\textwidth}
\centering
\begin{lstlisting}[basicstyle=\tiny\ttfamily,,keywordstyle=\ttfamily]
inspector = init_inspector(...);

// Three sets, edges, cells, and vertices
E = set(inspector, "edges", core_edges, boundary_edges, nonexec_edges, ...);
C = set(inspector, "cells", core_cells, boundary_cells, nonexec_cells, ...);
V = set(inspector, "verts", core_verts, boundary_verts, nonexec_verts, ...);

// Two maps, from edges to vertices and from cells to vertices
e2vMap = map(inspector, E, V, edges2verts, ...);
c2vMap = map(inspector, C, V, cells2verts, ...);

// The loop chain comprises three loops
// Each loop has some descriptors
loop(inspector, E, { $\perp$, "r"}, {e2vMap, "i"});
loop(inspector, C, { $\perp$, "r"}, {c2vMap, "i"});
loop(inspector, E, { $\perp$, "w"}, {e2vMap, "r"});

// Now can run the inspector
return inspection(mode, inspector, tile_size, ...);
\end{lstlisting}
\subcaption{Loop chain for the example program.}
\label{code:tiling-inspector}
\end{subfigure}

\end{CenteredBox}
\caption{On the left, a ``toy'' program used as running example in Section~\ref{sec:examples} to illustrate the loop chain abstraction and show how the sparse tiling algorithms (inspection, execution) work. Note that all parameters passed to the kernels are pointers. On the right, a code snippet showing the loop chain corresponding to the program on the left. The syntax is very close to the actual API of SLOPE, the sparse tiling library that we have implemented, described in Section~\ref{sec:implementation}.}
\end{figure}

\section{Loop Chain, Inspection and Execution Examples}
\label{sec:examples}

Using the example in Figure~\ref{code:tiling-runningexample}, we describe the actions performed by our sparse tiling inspector. The inspector takes as input the loop chain illustrated in Figure~\ref{code:tiling-inspector}. We show two variants, for shared-memory and distributed-memory parallelism. The value of the variable {\tt mode} in Figure~\ref{code:tiling-inspector} determines the variant to be executed.

\subsection{Overview}
\begin{figure}
\centering
\includegraphics[scale=0.43]{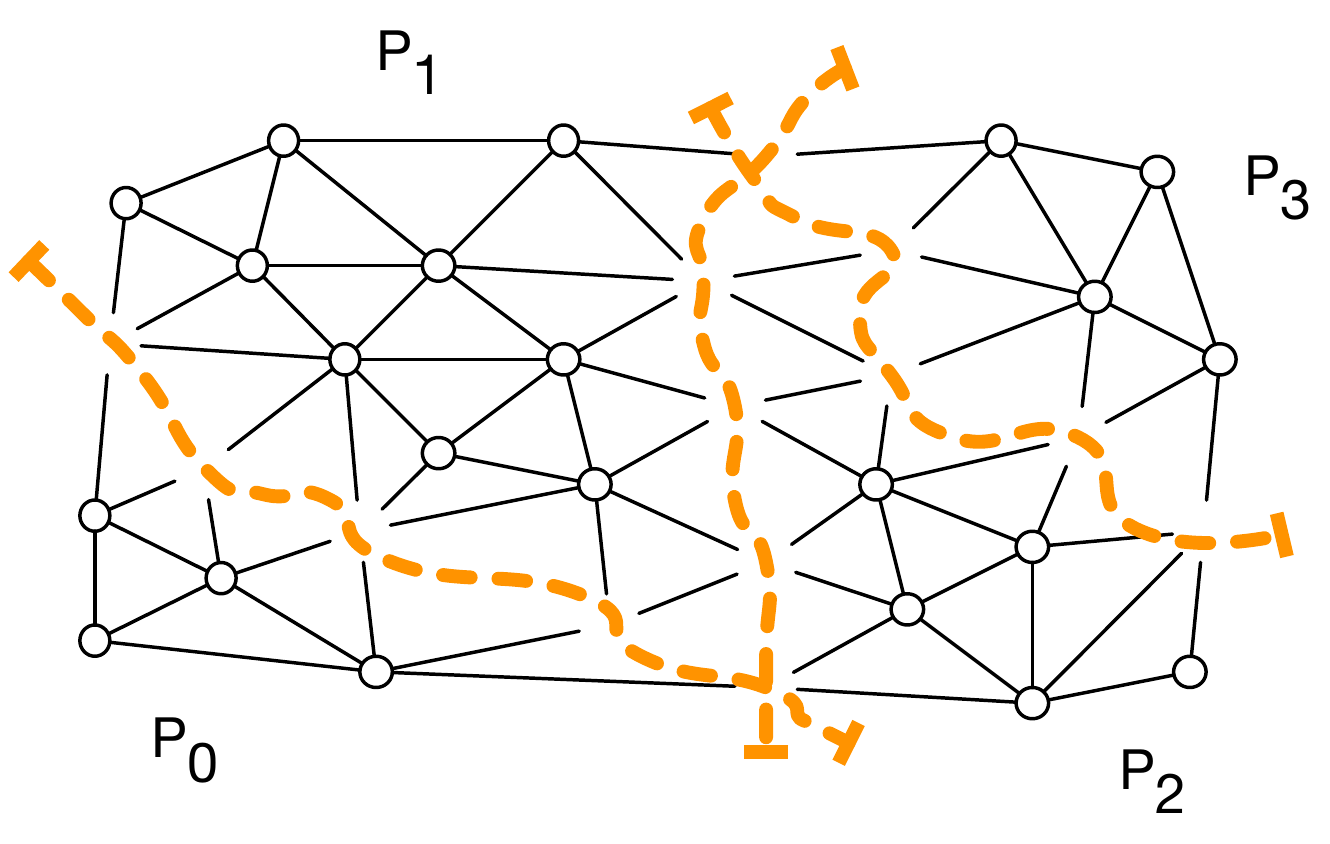}
\caption{Partitioning of the seed loop over edges. The vertices are illustrated to make the connectivity of the mesh clear, although they do not belong to any partition yet.}
\label{fig:st-initial-part-sm}
\end{figure}

The inspector starts with partitioning the iteration space of a {\it seed loop}, for example $L_0$. Partitions are used to initialize tiles: the iterations of $L_0$ falling in $P_i$ -- or, in other words, the edges in partition $P_i$ -- are assigned to the tile $T_i$. Figure~\ref{fig:st-initial-part-sm} displays the situation after the initial partitioning of $L_0$ for a given input mesh. There are four partitions, two of which ($P_0$ and $P_3$) are not connected through any edge or cell. These four partitions correspond to four tiles, $[T_0,\ T_1,\ T_2,\ T_3]$, with $P_i = T_i$.

As detailed in the next two sections, the inspection proceeds by populating $T_i$ with iterations from $L_1$ and $L_2$. The challenge of this task is guaranteeing that all data dependencies -- read after write, write after read, write after write -- are honored. The output of the inspector is eventually passed to the executor. The inspection carries sufficient information for computing sets of tiles in parallel. $T_i$ is always executed by a single thread/process and the execution is atomic; that is, it does not require communication with other threads/processes. When executing $T_i$, first all iterations from $L_0$ are executed, then all iterations from $L_1$ and finally those from $L_2$.

\subsection{Inspection for Shared-Memory Parallelism}
\label{sec:examples:shm}
\begin{figure}[t]

\begin{subfigure}{0.48\textwidth}
\centering
\includegraphics[scale=0.33]{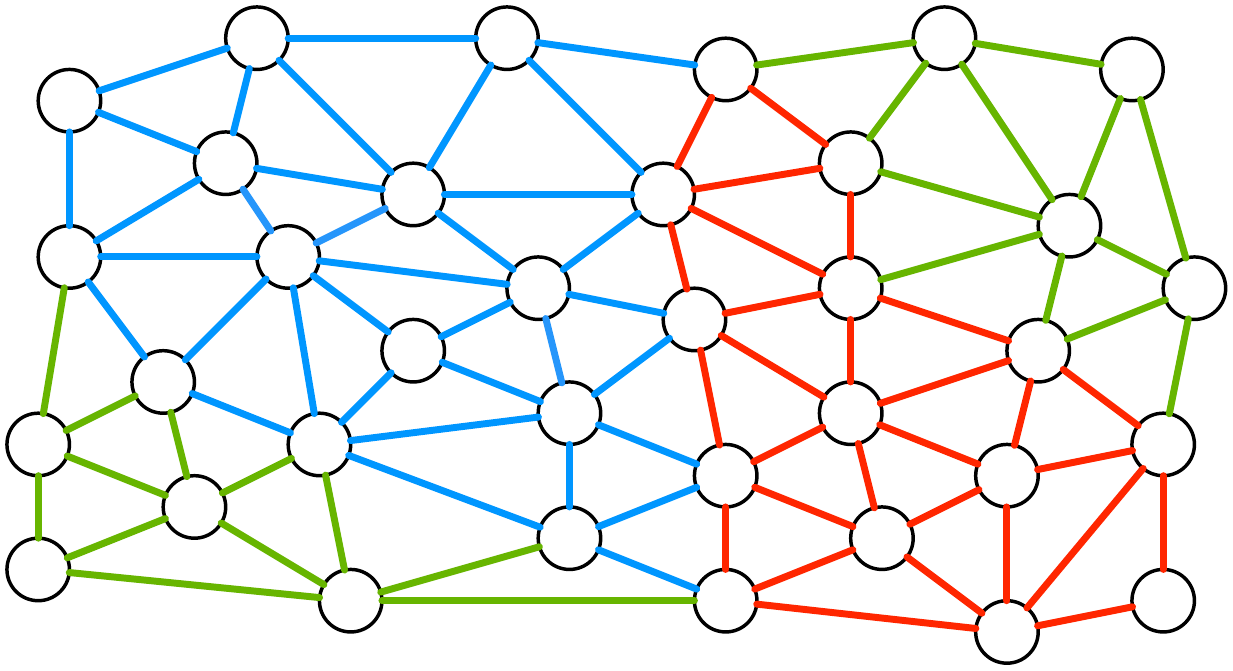}
\caption{A snapshot of the mesh after tiling $L_0$.}
\label{fig:st-loop-0}
\end{subfigure}%
~
\begin{subfigure}{0.48\textwidth}
\centering
\includegraphics[scale=0.33]{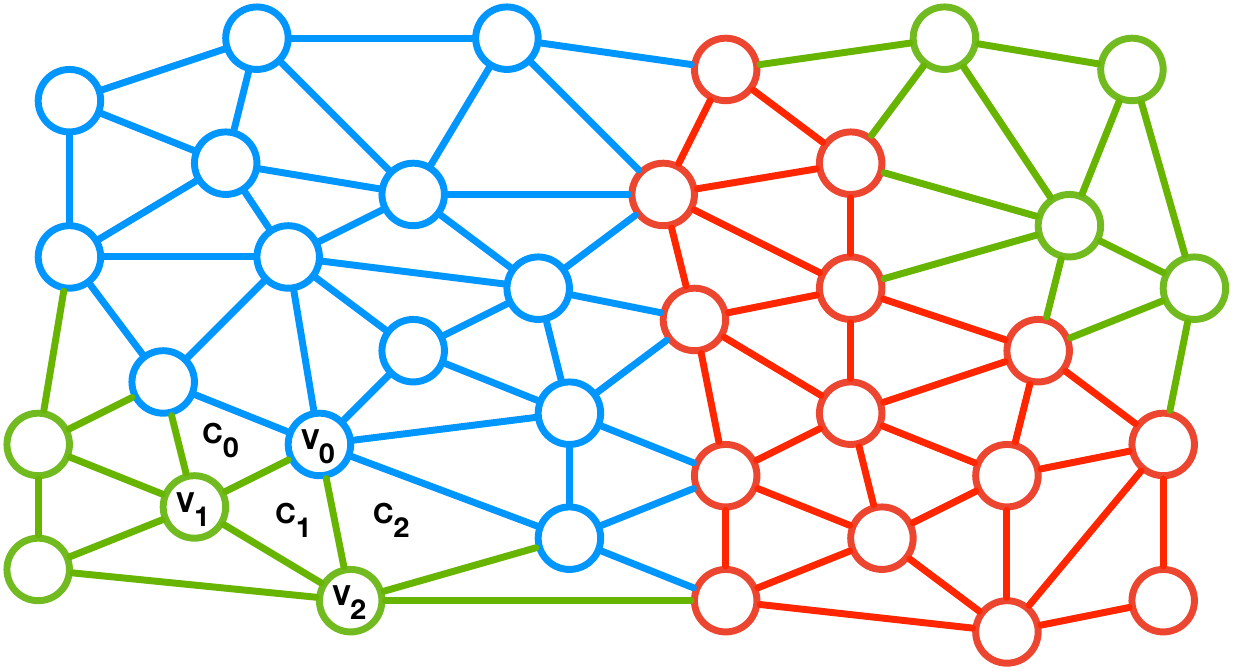}
\caption{The vertices are written by $L_0$, so a projection must be computed before tiling $L_1$. Here, the projection is represented as colored vertices.}
\label{fig:st-loop-0-proj}
\end{subfigure}
~\\~\\
\begin{subfigure}{0.48\textwidth}
\centering
\includegraphics[scale=0.33]{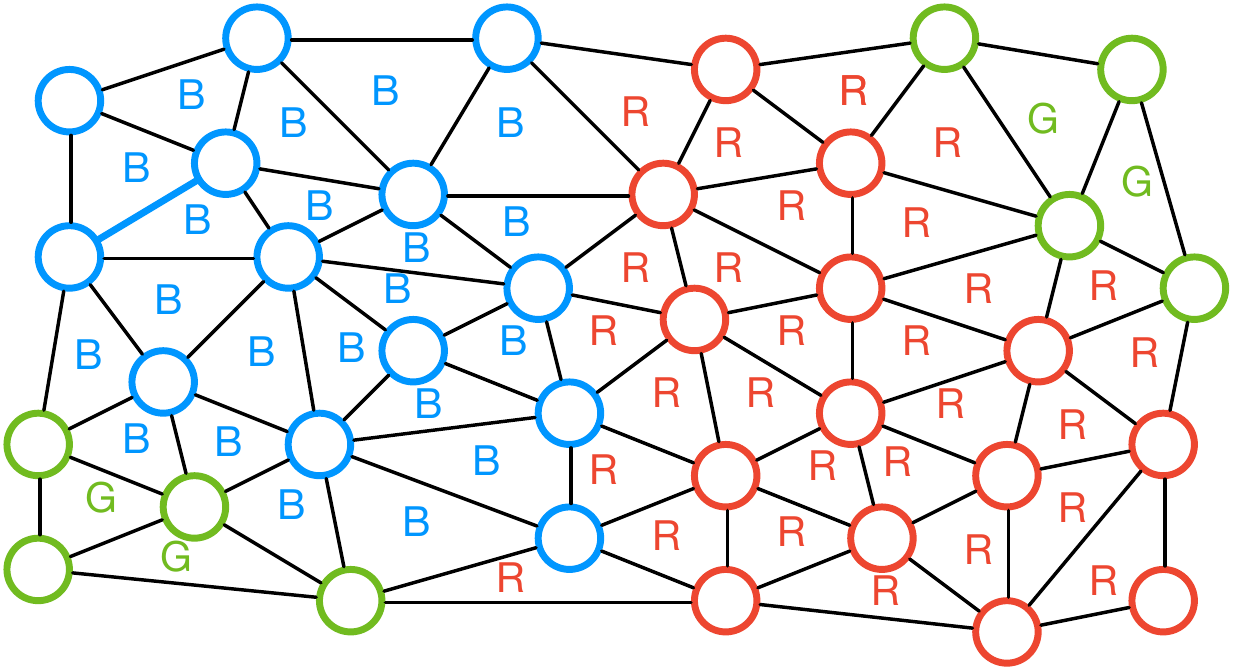}
\caption{A snapshot of the mesh after tiling $L_1$.}
\label{fig:st-loop-1}
\end{subfigure}
~
\begin{subfigure}{0.48\textwidth}
\centering
\includegraphics[scale=0.33]{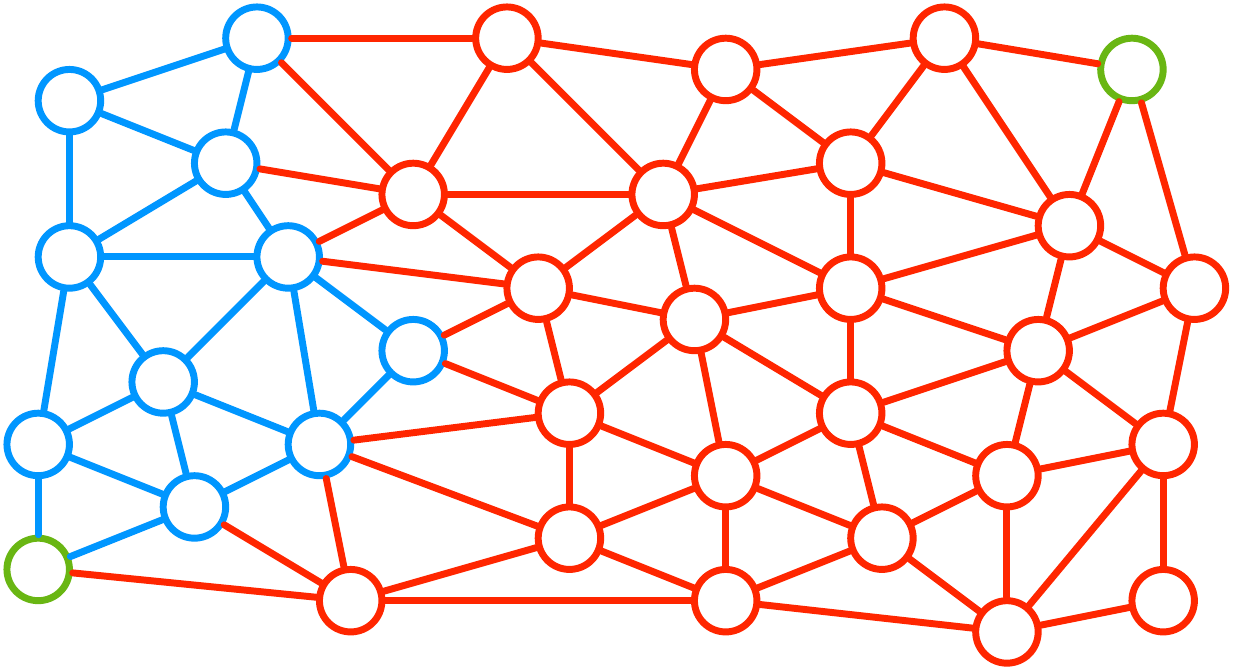}
\caption{A snapshot of the mesh after tiling $L_2$.}
\label{fig:st-loop-2}
\end{subfigure}

\caption{Four passes of the inspection algorithm for shared-memory parallelism.}
\label{fig:inspection-example}
\end{figure}

Similarly to OP2, to achieve shared-memory parallelism we use coloring. Two tiles that are given the same color can be executed in parallel by different threads. Two tiles can have the same color if they are not connected, because this ensures the absence of race conditions through indirect memory accesses during parallel execution. In the example we can use three colors: red (R), green (G), and blue (B). $T_0$ and $T_3$ are not connected, so they are assigned the same color. The colored tiles are shown in Figure~\ref{fig:st-loop-0}. In the following, with the notation $T_i^c$ we indicate that the $i$-th tile has color $c$.

To populate $[T_0^G,\ T_1^B,\ T_2^R,\ T_3^G]$ with iterations from $L_1$ and $L_2$, we first have to establish a total ordering for the execution of partitions with different colors. Here, we assume the following order: green (G), blue (B), and red (R). This implies, for instance, that {\it all iterations} assigned to $T_1^B$ must be executed {\it before all iterations} assigned to $T_2^R$. By ``all iterations'' we mean the iterations from $L_0$ (determined by the seed partitioning) as well as the iterations that will later be assigned from tiling $L_1$ and $L_2$. We assign integer positive numbers to colors to reflect their ordering, where a smaller number means higher execution priority. We can assign, for example, 0 to green, 1 to blue, and 2 to red.

To schedule the iterations of $L_1$ to $[T_0^G,\ T_1^B,\ T_2^R,\ T_3^G]$, we need to compute a {\it projection} for any write or local reduction performed by $L_0$. The projection required by $L_0$ is a function $\phi : V \rightarrow \mathbb{T}$ mapping the vertices in $V$ -- as indirectly incremented during the execution of $L_0$, see Figure~\ref{code:tiling-runningexample} -- to a tile $T_i^c \in \mathbb{T}$. Consider the vertex $v_0$ in Figure~\ref{fig:st-loop-0-proj}. $v_0$ has 7 incident edges, 2 of which belong to $T_0^G$, while the remaining 5 to $T_1^B$. Since we established that $G \prec B$, $v_0$ can only be read after $T_1^B$ has finished executing the iterations from $L_0$ (i.e., the 5 incident blue edges). We express this condition by setting $\phi(v_0) = T_1^B$. Observe that we can compute $\phi$ by iterating over $V$ and, for each vertex, applying the maximum function ($\operatorname{MAX}$) to the color of the adjacent edges.

We now use $\phi$ to schedule $L_1$, a loop over cells, to the tiles. Consider again $v_0$ and the adjacent cells $[c_0,\ c_1,\ c_2]$ in Figure~\ref{fig:st-loop-0-proj}. These three cells have in common the fact that they are adjacent to both green and blue vertices. For $c_1$, and similarly for the other cells, we compute $\operatorname{MAX}(\phi(v_0),\ \phi(v_1),\ \phi(v_2)) = \operatorname{MAX}(B, G, G) = B = 1$. This establishes that $c_1$ must be assigned to $T_1^B$, because otherwise ($c_1$ assigned instead to $T_0^G$) a read to $v_0$ would occur before the last increment from $T_1^B$ took place. Indeed, we recall that the execution order, for correctness, must be ``all iterations from $[L_0, L_1, L_2]$ in the green tiles before all iterations from $[L_0, L_1, L_2]$ in the blue tiles''. The scheduling of $L_1$ to tiles is displayed in Figure~\ref{fig:st-loop-1}.

To schedule $L_2$ to $[T_0^G,\ T_1^B,\ T_2^R,\ T_3^G]$ we employ a similar process. Vertices are again written by $L_1$, so a new projection over $V$ will be necessary. Figure~\ref{fig:st-loop-2} shows the output of this last phase.

\paragraph{Conflicting Colors}

\begin{figure}[hbtp]
\centering
\begin{subfigure}[b]{0.33\textwidth}
\includegraphics[width=\textwidth]{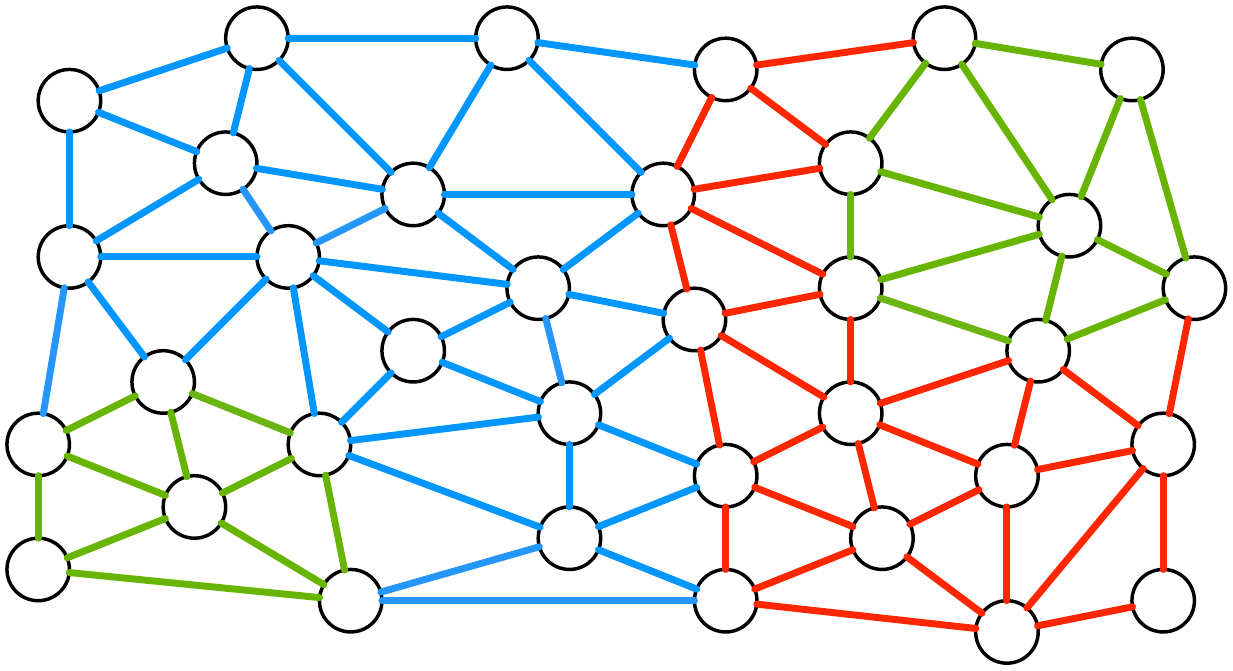}
\caption{After tiling $L_0$}
\label{fig:st-conflicts-a}
\end{subfigure}%
~ 
\begin{subfigure}[b]{0.33\textwidth}
\centering
\includegraphics[width=\textwidth]{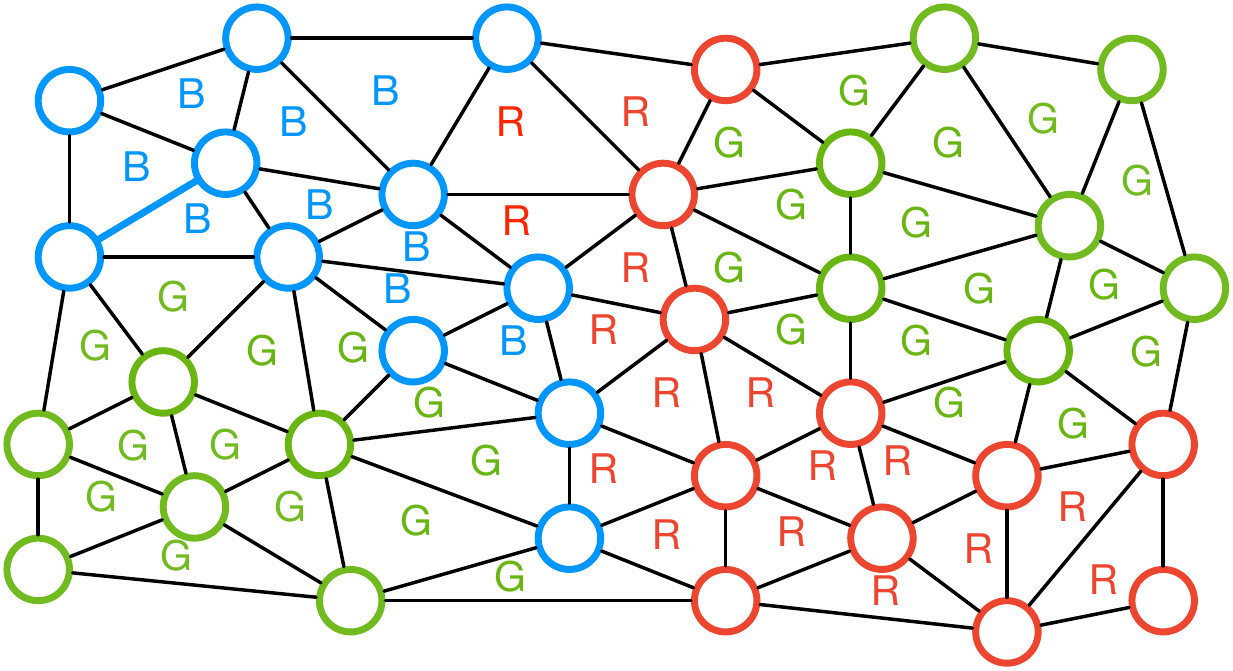}
\caption{After tiling $L_1$}
\label{fig:st-conflicts-b}
\end{subfigure}%
~
\begin{subfigure}[b]{0.34\textwidth}
\centering
\includegraphics[width=\textwidth]{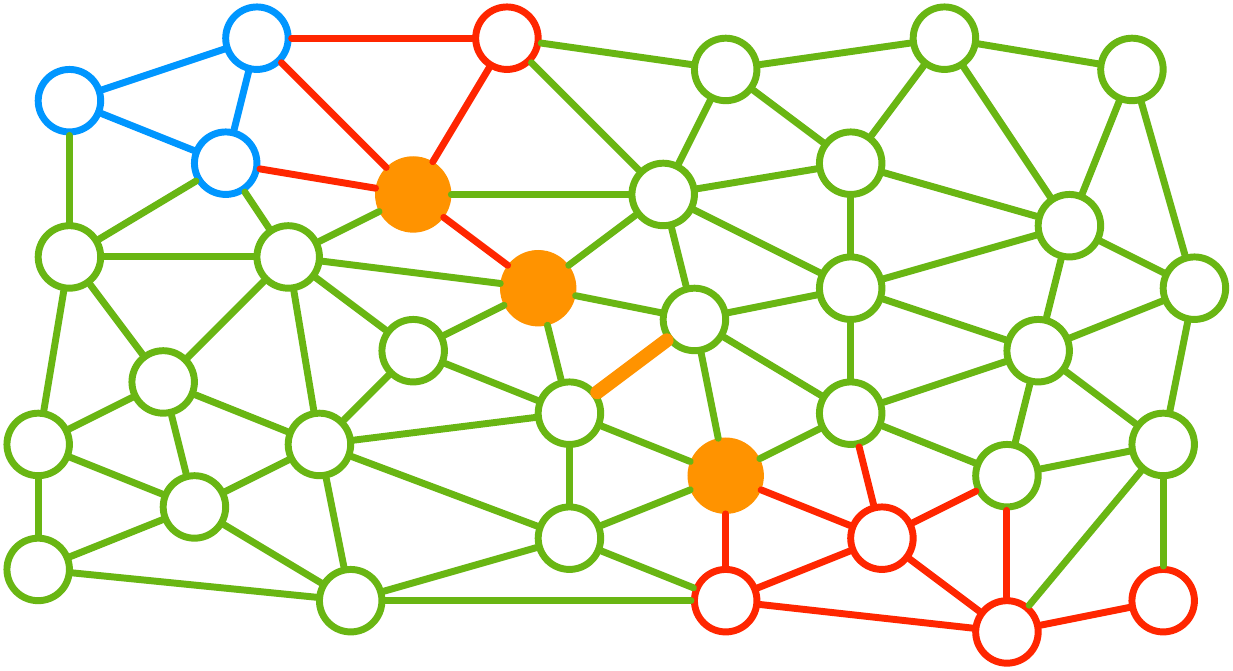}
\caption{After tiling $L_2$}
\label{fig:st-conflicts-c}
\end{subfigure}%

\caption{Tiling the program in Figure~\ref{code:tiling-runningexample} for shared-memory parallelism can lead to conflicts. Here, the two green tiles eventually become adjacent, creating race conditions.}
\label{fig:st-conflicts}
\end{figure}

It is worth noting how $T_2^R$ ``consumed'' the frontier elements of all other tiles every time a new loop was scheduled. Tiling a loop chain consisting of $k$ loops has the effect of expanding the frontier of a tile of at most $k$ vertices. With this in mind, we re-inspect the loop chain of the running example, although this time employing a different execution order -- blue (B), red (R), and green (G) -- and a different seed partitioning. Figure~\ref{fig:st-conflicts} shows that, by applying the same procedure described in this section, $T_0^G$ and $T_3^G$ will eventually become adjacent. This violates the precondition that {\it tiles can be given the same color, and thus run in parallel, as long as they are not adjacent}. Race conditions during the execution of iterations belonging to $L_2$ are now possible. This problem will be solved in Section~\ref{sec:algorithm}.

\subsection{Inspection for Distributed-Memory Parallelism}
\label{sec:examples:execution}
In the case of distributed-memory parallelism, the mesh is partitioned and distributed to a set of processes. Neighboring processes typically exchange (MPI) messages before executing a loop $L_j$. A message includes all ``dirty'' dataset values required by $L_j$ modified by any $L_k$, with $L_k \prec L_j$. In the running example, $L_0$ writes to vertices, so a subset of values associated with border vertices must be communicated prior to the execution of $L_1$. To apply sparse tiling, the idea is to push all communications to the beginning of the loop chain: as we shall see, this increases the amount of data to be communicated at a time, but also reduces the number of synchronizations (only 1 synchronization between each pair of neighboring processes per loop chain execution).

From Section~\ref{sec:tiling:lc} it is known that, in a loop chain, a set is logically split into three regions, {\it core}, {\it boundary}, and {\it non-exec}. The boundary tiles, which originate from the seed partitioning of the boundary region, will include all iterations that cannot be executed until the communications have terminated. The procedure described for shared-memory parallelism -- now performed individually by each process on a partition of the input mesh -- is modified as follows:
\begin{enumerate}
\item The core region of the seed loop $L_0$ is partitioned into tiles. Unless aiming for a mixed distributed/shared-memory scheme, there is no need to assign identical colors to unconnected tiles, as a process will execute its own tiles sequentially. Colors are assigned increasingly, with $T_i$ given color $i$. As long as tiles with contiguous ID are also physically contiguous in the mesh, this assignment retains memory access locality when ``jumping'' from executing $T_i$ to $T_{i+1}$.
\item The same process is applied to the boundary region. Thus, a situation in which a tile includes iterations from both the core and the boundary regions is prevented by construction. Further, all tiles within the boundary region are assigned colors higher than those used for the core tiles. This constrains the execution order: no boundary tiles will be executed until all core tiles are computed.
\item We map the whole non-exec region of $L_0$ to a single special tile, $T_{ne}$. This tile has the highest color and will actually never be executed. 
\end{enumerate}

\begin{figure}[h]
\centering
\includegraphics[scale=0.45]{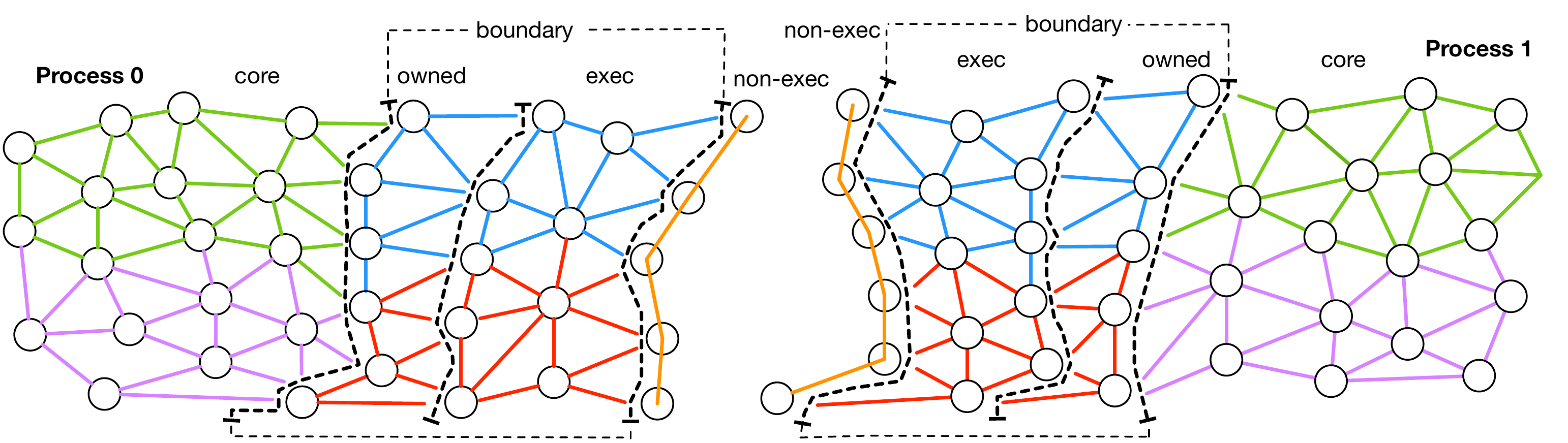}
\caption{A snapshot of the two mesh partitions on {\tt Process 0} and {\tt Process 1} after inspecting the seed loop $L_0$ for distributed-memory parallelism. On each process, there are five tiles in total: two in the core region (green and violet), two in the boundary region (red and light blue), and $T_{ne}$. The boundary tiles can safely cross the owned and exec sub-regions (i.e., the local iterations and the iterations to be redundantly computed, respectively). However, no tile can include iterations from both the core and the boundary regions. }
\label{fig:st-mpi-init}
\end{figure}

\begin{figure}[h]
\centering
\includegraphics[scale=0.45]{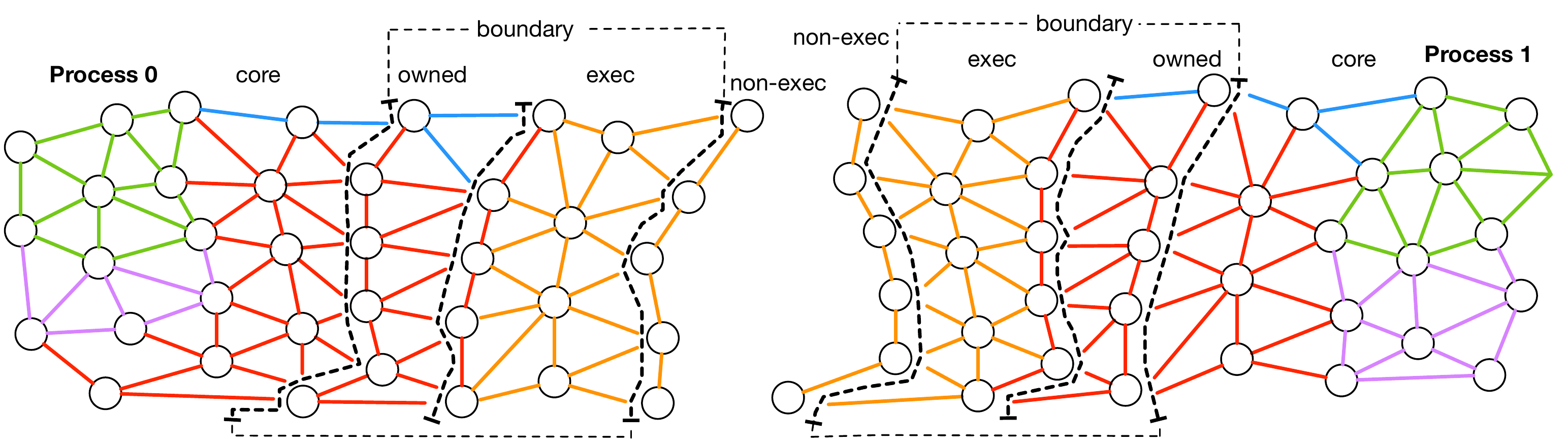}
\caption{A snapshot of the two mesh partitions on {\tt Process 0} and {\tt Process 1} at the end of the inspection for distributed-memory parallelism. $T_{ne}$ expands over the boundary region, which minimizes the amount of redundant computation to be performed. At the end of the execution phase, the orange edges will contain ``dirty  values'', but correctness is not affected as the exec region only includes off-process data. The boundary tiles expand over the core region: this is essential for correctness since none of the red and blue entities from $[L_0,\ L_1,\ L_2]$ can be executed until the MPI communications have terminated.}
\label{fig:st-mpi-growth}
\end{figure}

From this point on, the inspection proceeds as in the case of shared-memory parallelism. The application of the $\operatorname{MAX}$ function when scheduling $L_1$ and $L_2$ makes higher color tiles (i.e., those having lower priority) ``expand over'' lower color ones. 

In Figure~\ref{fig:st-mpi-init}, a mesh is partitioned over two processes and a possible seed partitioning and tiling of $L_0$ illustrated. We observe that the two boundary tiles (the red and light blue ones) will expand over the core tiles as $L_1$ and $L_2$ are tiled, which eventually results in the scheduling illustrated in Figure~\ref{fig:st-mpi-growth}. Roughly speaking, if a loop chain consists of $n$ loops and, on each process, $n-1$ extra layers of iterations are provided (the exec regions in Figure~\ref{fig:st-mpi-init}), then all boundary tiles are correctly computed. 

The schedule produced by the inspector is subsequently used by the executor. On each process, the executor starts with triggering the MPI communications required for the computation of boundary tiles. All core tiles are then computed, since no data from the boundary region is necessary. Hence, computation is overlapped with communication. As all core tiles are computed and the MPI communications terminated, the boundary tiles can finally be computed.

\paragraph{Efficiency considerations}
The underlying hypothesis is that the increase in data locality will outweigh the overhead induced by the redundant computation and by the bigger volume of data exchanged. This is motivated by several facts: (i) the loops being memory-bound;  (ii) the core region being much larger than the boundary region; (iii) the amount of redundant computation being minimized through the special tile $T_{ne}$, which progressively expands over the boundary region, thus avoiding unnecessary calculations.

\section{Data Dependency Analysis}
\label{sec:data-dep-analysis}
The loop chain abstraction, described in Section~\ref{sec:tiling:lc}, provides the information to construct an inspector capable of analyzing data dependencies and thus build a legal sparse tiling schedule. The dependencies between two different loops may be of type flow (read after write), anti (write after read), or output (write after write). Further, there may be ``reduction dependencies'' between iterations of the same loop. 

\subsection{Cross-Loop Dependences}
Assume that loop $L_x$, having iteration space $S_x$, precedes loop $L_y$, having iteration space $S_y$, in the loop chain. Let $e_x$ be a generic iteration in $S_x$. Let $M$ be a map of arity $a$ between two iteration spaces. Let $\texttt{mode} \in \lbrace w, r, i\rbrace$ indicate whether an iteration is written, read, or incremented. We represent direct and indirect accesses as follows.
\begin{description}
\item[Direct access] $\perp_{S_x}^{\mathrm{mode}}(e_x) \rightarrow \lbrace \lbrace e_x \rbrace, \emptyset \rbrace$. In particular, if $\perp_{S_x}^{\mathrm{mode}}(e_x) = \emptyset$, then no direct write/read/increment is performed by $e_x$ when computing $L_x$.
\item[Indirect access] $M_{S_x \rightarrow S_y}^{\mathrm{mode}}(e_x) \rightarrow \lbrace \lbrace e_{y_0}, ..., e_{y_{a-1}} \rbrace, \emptyset \rbrace$. As per direct accesses, $M_{S_x \rightarrow S_y}^{\mathrm{mode}}(e_x) = \emptyset$ means that $e_x$ does not indirectly write/read/increment $S_y$ when computing $L_x$.
\end{description}

A direct access is a special case of indirect access when $y=x$ and $M_{S_x \rightarrow S_y}$ is the identity mapping. However, we here keep the distinction between the two types of access explicit due to their relevance in the sparse tiling algorithms, as explained in Section~\ref{sec:algorithm}.

By considering pairs of points $(e_x, e_y)$ in the iteration spaces of the two loops $L_x$ and $L_y$, namely $e_x \in S_x$ and $e_y \in S_y$, we can enumerate all possible dependences. For brevity, we do not distinguish between increments and writes; we also assume that at least one of the two loops accesses data indirectly. Let $S_z$ be a generic iteration space in the loop chain. Hence, the flow dependences are:
\begin{align*}
&	\{ e_x \rightarrow e_y \; |
    \underbrace{(\perp_{S_x}^{w}(e_x) \cap M_{S_y \rightarrow S_x}^{r}(e_y))}_\text{\ding{192} direct w, indirect r} \cup
    \underbrace{(M_{S_x \rightarrow S_y}^{w}(e_x) \cap \perp_{S_y}^{r}(e_y))}_\text{\ding{193} indirect w, direct r} \cup
    \underbrace{(M_{S_x\rightarrow S_z}^{w}(e_x) \cap M_{S_y \rightarrow S_z}^{r}(e_y))}_\text{\ding{194} indirect w, indirect r} \ne \emptyset
	\}.
\end{align*}
In essence, there is a flow dependence between two iterations from different loops when one of those iterations writes to an element and the other iteration reads from the same element, directly or indirectly. To capture all these flow dependences, the inspection algorithm builds {\it projections} from one loop to another. We saw an example in Section~\ref{sec:examples:shm}: the loop over cells ($S_x$) performed an indirect increment to a dataset associated with vertices ($S_z$), which was then read by the subsequent loop over edges ($S_y$). Such flow dependence was of type \ding{194} (see definition above). For each vertex $e_z \in S_z$, the projection (illustrated in Figure~\ref{fig:st-loop-0-proj}) captured the {\it last tile} indirectly writing to (incrementing) $e_z$, exploiting the color (i.e., the scheduling priority) of the source iterations. A flow dependence of type \ding{192} would be even simpler to deal with, as it would not require the use of the indirect map $M_{S_x \rightarrow S_z}$ to update the color of the iterations.

Likewise, we can enumerate the anti and output dependences:
\begin{align*}
&	\{ e_x \rightarrow e_y \; |
    (\perp_{S_x}^{r}(e_x) \cap M_{S_y \rightarrow S_x}^{w}(e_y)) \cup
    (M_{S_x \rightarrow S_y}^{r}(e_x) \cap \perp_{S_y}^{w}(e_y)) \cup
	(M_{S_x\rightarrow S_z}^{r}(e_x) \cap M_{S_y \rightarrow S_z}^{w}(e_y)) \ne \emptyset \}.\\
&	\{ e_x \rightarrow e_y \; |
    (\perp_{S_x}^{w}(e_x) \cap M_{S_y \rightarrow S_x}^{w}(e_y)) \cup
    (M_{S_x \rightarrow S_y}^{w}(e_x) \cap \perp_{S_y}^{w}(e_y)) \cup
	(M_{S_x\rightarrow S_z}^{w}(e_x) \cap M_{S_y \rightarrow S_z}^{w}(e_y)) \ne \emptyset \}.
\end{align*}
Projections for this type of dependences are built analogously to that described above.

The inspection algorithm building projections for all flow, anti, and output dependences is provided in Algorithm~\ref{algo:st-projection} and discussed in Section~\ref{sec:inspector:proj}. How the inspector leverages data dependence analysis (i.e., projections) to schedule iterations to tiles (i.e., the tiling function) is formalized in Algorithm~\ref{algo:st-tiling} and commented in Section~\ref{sec:inspector:tiling}.

\subsection{Intra-Loop Dependences}
There also are local reductions, or ``reduction dependencies'', between two or more iterations of the same loop when those iterations increment the same location(s); that is, when they read, modify with a commutative and associative operator, and write to the same location(s). The reduction dependencies in $L_x$ are:
\begin{align*}
\{ e_{x_1} \rightarrow e_{x_2} \; |  M_{S_x\rightarrow S_z}^{i}(e_{x_1}) \cap M_{S_x \rightarrow S_z}^{i}(e_{x_2}) \ne \emptyset \}.
\end{align*}
A reduction dependency between two iterations within the same loop indicates that those two iterations must be executed atomically with respect to each other. As we explained in Section~\ref{sec:examples:shm}, the inspection algorithm uses coloring to ensure atomic increments.

\section{Algorithms}
\label{sec:algorithm}

\begin{table}[h]
\tiny
\centering
\begin{tabulary}{1.0\columnwidth}{C|C}
\hline
Symbol & Meaning \\
\hlineB{4}
$\mathbb{L}$ & The loop chain \\
$L_j$ & The $j$-th loop in $\mathbb{L}$ \\
$\mathrm{seed}$ & The index of the seed loop\\
$S_j$ & The iteration space of $L_j$ \\
$S_j^{c}$, $S_j^{b}$, $S_j^{\mathrm{ne}}$ & The core, boundary, and non-exec regions of $S_j$ \\ 
$D$ & A descriptor of a loop \\
$\mathbb{T}$ & The collection of tiles \\
$\mathbb{T}[i]$ & Accessing the $i$-th tile, or $T_i$ \\
$T_i^{c}$, $T_i^{b}$, $T_i^{\mathrm{ne}}$ & The core, boundary, and non-exec regions of $T_i$\\
$\phi_S$ & A projection $\phi_S : S \rightarrow \mathbb{T}$ \\
$\Phi$ & The collection of projections \\
$\sigma_j$ & A tiling function $\sigma_j : S_j \rightarrow \mathbb{T}$ for $L_j$ \\
$\mathrm{ts}$ & The seed tile size \\
$C$ & The matrix of conflicting colors \\
{\tt Ax:y} & Algorithm x, line y \\
\hline
\end{tabulary}
\caption{Summary of the notation used throughout Section~\ref{sec:algorithm}.}
\label{table:st-summary-notation}
\end{table}

The pseudo-code for the sparse tiling inspector is shown in Algorithm~\ref{algo:st-inspector}. Given a loop chain and a seed tile size, the algorithm produces a schedule suitable for mixed distributed/shared-memory parallelism. This schedule -- in essence, a set of populated tiles -- is used by the executor to perform the sparse-tiled computation. The executor pseudo-code is displayed in Algorithm~\ref{algo:st-executor}. The next two sections elaborate on the main steps of these algorithms. The notation is summarized in Table~\ref{table:st-summary-notation}; the syntax {\tt Ax:y} is a shortcut for ``Algorithm x, line y''. The implementation is discussed in Section~\ref{sec:implementation}.

\begin{minipage}[t]{6.7cm}
\IncMargin{1em}
\begin{algorithm}[H]
\SetAlgoLined
\caption{The inspection algorithm}\label{algo:st-inspector}
\SetKwData{SeedMap}{seed$\_$map}
\SetKwData{Conflicts}{conflicts}
\SetKwData{C}{C}
\SetKwFunction{AFC}{add$\_$fake$\_$connection}
\SetKwFunction{IFC}{has$\_$conflicts}
\SetKwFunction{CLM}{compute$\_$local$\_$maps}
\SetKwFunction{Color}{color}
\SetKwFunction{Partition}{partition}
\SetKwFunction{FindMap}{find$\_$map}
\SetKwFunction{Project}{project}
\SetKwFunction{Assign}{assign}
\SetKwFunction{Tile}{tile}
\tiny

\kwInput{The loop chain $\mathbb{L} = [L_0,\ L_1,\ ...,\ L_{n-1}]$, a seed tile size $\mathrm{ts}$}
\kwOutput{A collection of tiles $\mathbb{T}$, populated with iterations from $\mathbb{L}$}

$\mathrm{seed} \gets 0$, $\Phi \gets \emptyset$, $\C \gets \perp$\;  \label{algo:insp-empty-projs}
$\sigma_{\mathrm{seed}}$, $\mathbb{T} \gets$ \Partition{$S_{\mathrm{seed}}$, $\mathrm{ts}$}\; \label{algo:insp-partition}
\SeedMap $\gets$ \FindMap{$S_{\mathrm{seed}}$, $\mathbb{L}$}\;
\Do{\Conflicts}{
  \Conflicts $\gets$ \False\;
  \Color{$\mathbb{T}$, \SeedMap}\; \label{algo:insp-color}
  \For{$j=1$ \KwTo $n-1$}{ \label{algo:st-tiling-loop}
    \Project{$L_{j-1}$, $\sigma_{j-1}$, $\Phi$, \C}\;
    $\sigma_j \gets$ \Tile{$L_j$, $\Phi$}\;
    \Assign{$\sigma_j$, $\mathbb{T}$}\; \label{algo:insp-assign}
  }
  \If{\IFC{\C}}{
    \Conflicts $\gets$ \True\;
    \AFC{\SeedMap, \C}\;
  }
}
\CLM{$\mathbb{T}$}\;
\Return{$\mathbb{T}$}
 \end{algorithm}
\end{minipage}%
\begin{minipage}[t]{6.7cm}
\IncMargin{1em}
\begin{algorithm}[H]
\caption{The executor algorithm}\label{algo:st-executor}
\SetKwData{Color}{color}
\SetKwData{T}{$T$}
\SetKwFunction{SMC}{start$\_$MPI$\_$comm}
\SetKwFunction{EMC}{end$\_$MPI$\_$comm}
\SetKwFunction{GTBR}{group$\_$tiles$\_$by$\_$region}
\SetKwFunction{ET}{execute$\_$tile}
\tiny

\kwInput{A collection of tiles $\mathbb{T}$}
\KwResult{Execute the loop chain}

$\mathbb{T}^{c}$, $\mathbb{T}^{b}$ $\gets$ \GTBR{$\mathbb{T}$}\;
\nonl ~\\
\SMC{}\;
\nonl ~\\
\ForEach{\Color}{
  \ForEach{$\T \in \mathbb{T}^{c}$ s.t. \T .\Color $==$ \Color}{ \label{algo:st-executor:parallel1}
    \ET{\T}\;
  }
}
\nonl ~\\
\EMC{}\;
\nonl ~\\
\ForEach{\Color}{
  \ForEach{$\T \in \mathbb{T}^{b}$ s.t. \T .\Color $==$ \Color}{ \label{algo:st-executor:parallel2}
    \ET{\T}\;
  }
}
  \end{algorithm}
\end{minipage}

\subsection{Inspector}
\label{sec:inspector}

\subsubsection{Choice of the seed loop}
The seed loop $L_{\mathrm{seed}}$ is used to initialize the tiles. Theoretically, any loop in the chain can be chosen as seed. Supporting distributed-memory parallelism, however, is cumbersome if $L_{\mathrm{seed}} \neq L_0$. This is because more general schemes for partitioning and coloring would be needed to ensure that no iterations in any $S_j^{b}$ are assigned to a core tile. A limitation of our inspector algorithm in the case of distributed-memory parallelism is that it must be $L_{\mathrm{seed}} = L_0$. 

In the special case in which there is no need to distinguish between core and boundary tiles (because a program is executed on a single shared-memory system), $L_{\mathrm{seed}}$ can be chosen arbitrarily. If we however pick $L_{\mathrm{seed}}$ in the middle of the loop chain, that is $L_0 \prec ... \prec L_{\mathrm{seed}} \prec ... \prec L_{n-1}$, a mechanism for constructing tiles in the reverse direction (``backwards''), from $L_{\mathrm{seed}}$ towards $L_0$, is necessary. In~\cite{st-paper}, we propose two ``symmetric'' algorithms to solve this problem, \textit{forward tiling} and \textit{backward tiling}, with the latter using the $\operatorname{MIN}$ function in place of $\operatorname{MAX}$ when computing projections. For ease of exposition, and since in the fundamental case of distributed-memory parallelism we are imposing $L_{\mathrm{seed}} = L_0$, we here neglect this distinction\footnote{The algorithm implemented in SLOPE, the library presented in Section~\ref{sec:tiling:impl-slope}, supports backwards tiling for shared-memory parallelism.}. 

\subsubsection{Tiles initialization}
\label{sec:inspector:init}
Let $\mathrm{ts}$ be the user-specified seed tile size. The algorithm starts with partitioning $S_{\mathrm{seed}}^{c}$ into $m$ subsets $\lbrace P_0, P_1, ..., P_{m-1}\rbrace$ such that $|P_i| = ts$ (except possibly for $P_{m-1}$), $P_i \cap P_j = \emptyset$, and $\cup_{i = 0}^{m-1} P_i = S_{\mathrm{seed}}^{c}$. Among all possible legal partitionings, we choose the one that splits $S_{\mathrm{seed}}^c$ into blocks of $\mathrm{ts}$ contiguous iterations, with $P_0 = \lbrace 0, ..., ts-1\rbrace$, $P_1 = \lbrace ts, ..., 2 ts - 1\rbrace$, and so on. We analogously partition $S_{\mathrm{seed}}^{b}$ into $k$ subsets. We create $m+k+1$ tiles, one for each of these partitions and one extra tile for $S_{\mathrm{seed}}^{\mathrm{ne}}$, namely $\mathbb{T} = [T_0^c, ..., T_{m-1}^c, T_m^{b}, ..., T_{m+k-1}^b, T_{m+k}^{\mathrm{ne}}]$. At this point we have an assignment of iterations to tiles for $L_{\mathrm{seed}}$; that is, a tiling function $\sigma_{\mathrm{seed}} : S_{\mathrm{seed}} \rightarrow \mathbb{T}$. This initial partitioning phase occurs at {\tt A\ref{algo:st-inspector}:\ref{algo:insp-partition}}.

A tile $T_i$ has four fields, as summarized in Table~\ref{table:st-tile-structure}. 

\begin{itemize}
\item The {\em region} is used by the executor to schedule tiles in a given order. This field is set right after the partitioning of $L_{\mathrm{seed}}$, as a tile (by construction) exclusively belongs to $S_{\mathrm{seed}}^c$, $S_{\mathrm{seed}}^b$, or $S_{\mathrm{seed}}^{\mathrm{ne}}$.
\item The {\em iteration lists} contain the iterations in $\mathbb{L}$ that $T_i$ will have to execute. There is one {\em iteration list} for each $L_j \in \mathbb{L}$, indicated as $[T_i]_j$ . At this stage of the inspection we have $[T_i]_{\mathrm{seed}} = [T_i]_0 = P_i$, whereas still $[T_i]_j = \emptyset$ for $j=1,...,n-1$.
\item {\em Local maps} may be used for performance optimization by the executor in place of the global maps provided through the loop chain; this will be discussed in more detail in Section~\ref{sec:performance} and in the Supplementary Materials.
\item The {\em color} provides a tile with a scheduling priority. If shared-memory parallelism is requested, adjacent tiles are given different colors (the adjacency relation is determined through the maps available in $\mathbb{L}$). Otherwise, colors are assigned in increasing order (i.e., $T_i$ is given color $i$). The boundary tiles are always given colors higher than that of core tiles; the non-exec tile has the highest color. The assignment of colors is carried out by {\tt A\ref{algo:st-inspector}:\ref{algo:insp-color}}.
\end{itemize}

\begin{table}[h]
\tiny
\centering
\begin{tabulary}{1.0\columnwidth}{P{2.7cm} | P{8.5cm}}
\hline
Field & Possible values \\
\hlineB{4}
{\em region} & core, boundary, non-exec \\
{\em iterations lists} & one list of iterations $[T_i]_j$ for each $L_j \in \mathbb{L}$\\
{\em local maps} & one list of local maps for each $L_j \in \mathbb{L}$; one local map for each map used in $L_j$\\
{\em color} & an integer representing the execution priority \\ 
\hline
\end{tabulary}
\caption{The tile data structure.}
\label{table:st-tile-structure}
\end{table}

\subsubsection{The inspection loop}
The inspection loop, starting at {\tt A\ref{algo:st-inspector}:\ref{algo:st-tiling-loop}}, schedules the remaining $L_j \in \mathbb{L}$ by alternating dependence analysis and construction of tiling functions. The input is $\sigma_{\mathrm{seed}}$. As seen in the previous sections, a projection is a function $\phi_S : S \rightarrow \mathbb{T}$ that captures data dependences across loops. Initially, the projections set $\Phi$ is empty ({\tt A\ref{algo:st-inspector}:\ref{algo:insp-empty-projs}}). Once a new loop is tiled, $\Phi$ is updated by adding new projections or changing existing ones (see Section~\ref{sec:inspector:proj}). Using $\Phi$, a new tiling function $\sigma_j$ for $L_j$ is derived (see Section~\ref{sec:inspector:tiling}).

\subsubsection{Deriving a projection from a tiling function}
\label{sec:inspector:proj}
Algorithm~\ref{algo:st-projection} takes as input (the descriptors of) an $L_{j}$ and its tiling function $\sigma_{j} :S_j \rightarrow \mathbb{T}$ to update $\Phi$. The algorithm also updates the conflicts matrix $C \in \mathbb{N}^{m \times m}$, which indicates whether two tiles having the same color will become adjacent.

\begin{figure}[h]
\begin{CenteredBox}
\includegraphics[scale=0.40]{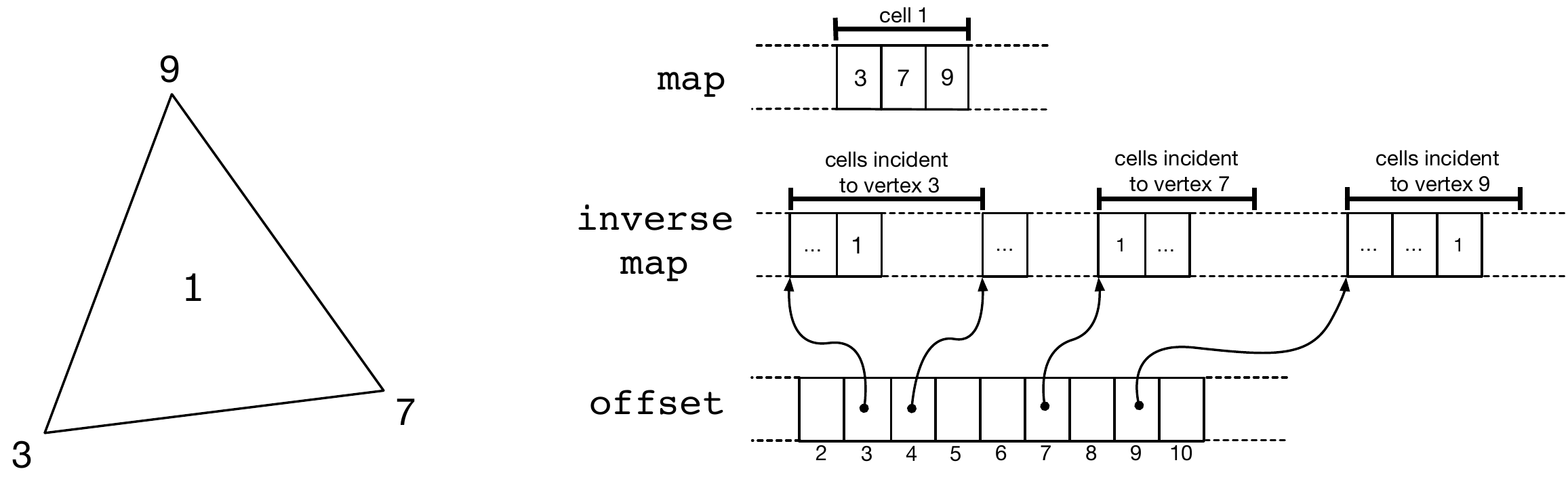}
\end{CenteredBox}
\caption{Representation of an inverse map. The original map shows that the triangular cell $1$ is adjacent to three vertices, namely $3$, $7$, and $9$. The inverse map associates vertices to cells. Since the mesh is unstructured, different vertices can be incident to a different number of cells. The array {\tt offset} provides the distance between two consecutive vertices in the inverse map. For instance, all entries in the inverse map between {\tt offset[3]} and {\tt offset[4]} are cells incident to vertex $3$.}
\label{fig:st-inverse-map}
\end{figure}

A new projection $\phi_{S}$ is needed if $S$ is written by $L_{j}$. As explained in Section~\ref{sec:data-dep-analysis}, $\phi_{S}$ carries the necessary information to tile a subsequent loop accessing $S$. Let us consider the non-trivial case in which $L_j$ writes indirectly to $S$ through a map $M : S_j \rightarrow S^a$. To compute $\phi_{S}$, we first determine the inverse map $M^{-1}$ ({\tt A\ref{algo:st-projection}:\ref{algo:st-projection-invert}}; an example is shown in Figure~\ref{fig:st-inverse-map}). Then, we iterate over all elements $e \in S$ and we set $\phi_{S}[e]$ to the last tile writing to $e$. This is accomplished by applying the $\operatorname{MAX}$ function over the color of the tiles accessing $e$ (see {\tt A\ref{algo:st-projection}:\ref{algo:st-projection-max}}), obtained through $M^{-1}$. This procedure was used, for example, to compute the projection in Figure~\ref{fig:st-loop-0-proj}.

\begin{minipage}[t]{6.7cm}
\IncMargin{1em}
\begin{algorithm}[H]
\caption{Projection of a tiled loop}\label{algo:st-projection}
\tiny
\SetKwData{Descriptors}{descriptors}
\SetKwData{Arity}{arity}
\SetKwData{T}{$T$}
\SetKwData{AT}{$T_{\mathrm{last}}$}
\SetKwData{MC}{max}
\SetKwData{IM}{$M^{-1}$}
\SetKwData{Map}{map}
\SetKwData{Mode}{mode}
\SetKwData{D}{D}
\SetKwData{C}{C}
\SetKwData{Values}{values}
\SetKwData{Offset}{offset}
\SetKwData{Color}{color}
\SetKwFunction{MapInvert}{map$\_$invert}
\SetKwFunction{Update}{update$\_$color$\_$conflicts}
\SetKwFunction{Unpack}{unpack}

\kwInput{A loop $L_j$, a tiling function $\sigma_j$, the projections set $\Phi$, the conflicts matrix \C}
\KwResult{Update $\Phi$ and \C}

\ForEach{\D $\in$ $L_j$.\Descriptors}{
  \eIf{\D .\Map $==$ $\perp$}{
    $\Phi[S_j] \gets \sigma_{j}$\;
  }{
    \IM $\gets$ \MapInvert{\D .\Map}\; \label{algo:st-projection-invert}
    $S$, $S_j$, \Values, \Offset $\gets$ \IM.\Unpack()\;
    $\phi_{S} \gets \perp$\; 
    \For{$e$ \In $S$}{ \label{algo:st-projection-parallel}
      \For{$k= \Offset[e]$ \KwTo $\Offset[e+1]$}{
        \AT = $\mathbb{T}[\Values[k]]$\;
        \MC $\gets$ MAX($\phi_{S}[e]$.\Color, \AT .\Color)\; \label{algo:st-projection-max}
        \If{\MC $\neq$ $\phi_{S}[e]$.\Color}{
          $\phi_{S}[e] \gets$ \AT\;
        }
      }
    }
    \Update{\C, $\mathbb{T}$, $\phi_{S}$}\;
    $\Phi[S] = \phi_{S}$\;
  }
}
  \end{algorithm}
\end{minipage}%
\begin{minipage}[t]{6.7cm}
\IncMargin{1em}
 \begin{algorithm}[H]
\caption{Building a tiling function}\label{algo:st-tiling}
\tiny

\SetKwData{Descriptors}{descriptors}
\SetKwData{Arity}{arity}
\SetKwData{AT}{$T$}
\SetKwData{MC}{max}
\SetKwData{Size}{size}
\SetKwData{Map}{map}
\SetKwData{D}{D}
\SetKwData{Values}{values}
\SetKwData{Color}{color}

\kwInput{A loop $L_{j}$, the projections set $\Phi$}
\kwOutput{The tiling function $\sigma_{j}$}

$\sigma_j \gets \perp$\;
\ForEach{\D $\in$ $L_j$.\Descriptors}{
  \eIf{\D .\Map $==$ $\perp$}{
    $\phi_{S_j} \gets \Phi[S_j]$\;
    \For{$e$ \In $S_j$}{
      \MC $\gets$ MAX($\sigma_j[e]$.\Color, $\phi_{S_j}[e]$.\Color)\;
      \If{\MC $\neq$ $\sigma_j[e]$.\Color}{
          $\sigma_j[e] \gets$ $\phi_{S_j}[e]$\;
        }
    }
  }{
    \Arity $\gets$ D.\Map .\Arity\;
    $\phi_{S} \gets \Phi[\D.\Map.S]$\;
    \For{$e$ \In $S_j$}{ \label{algo:st-tiling-parallel}
       $\sigma_j[e] \gets \AT_{\perp}$\;
      \For{$k=0$ \KwTo \Arity}{
        \AT $\gets \phi_{S}[\D.\Map .\Values[e*\Arity + k]]$\;
        \MC $\gets$ MAX($\sigma_j[e]$.\Color, \AT .\Color)\;
        \If{\MC $\neq$ $\sigma_j[e]$.\Color}{
          $\sigma_j[e] \gets$ \AT\;
        }
      }
    }
  }
}
\Return{$\sigma_j$}
  \end{algorithm}
\end{minipage}

\subsubsection{Deriving a tiling function from the available projections}
\label{sec:inspector:tiling}
Using $\Phi$, we compute $\sigma_j$ as described in Algorithm~\ref{algo:st-tiling}. The algorithm is similar to the projection of a tiled loop (e.g., maps are used to access the neighborhood of a target iteration). The main difference is that now the projections in $\Phi$ are used, rather than computed, to schedule iterations to tiles so that data dependences are honored. Finally, $\sigma_j$ is used to populate the iteration lists $[T_i]_j$, for all $T_i \in \mathbb{T}$ (see {\tt A\ref{algo:st-inspector}:\ref{algo:insp-assign}}).

\subsubsection{Detection of conflicts}
If $C$ indicates the presence of at least one conflict, say between $T_{i_1}$ and $T_{i_2}$, we add a ``fake connection'' between these two tiles and loop back to the coloring stage. $T_{i_1}$ and $T_{i_2}$ are now connected, so they will be assigned different colors. 

\subsubsection{On the history of the algorithm}
The first algorithm for generalized sparse tiling inspection was introduced in~\cite{st-paper}. In this section, a new, enhanced version of that algorithm has been presented. In essence, the major differences are: (i) support for distributed-memory parallelism; (ii) use of coloring instead of a task graph for tile scheduling; (iii) speculative inspection with backtracking if a coloring conflict is detected; (iv) use of sets, instead of datasets, for data dependency analysis; (v) use of inverse maps for parallelization of the projection and tiling routines; (vi) computation of local maps. Most of these changes contributed to reduce the inspection cost, as discussed later.

\subsection{Executor}
\label{sec:executor}

The sparse tiling executor is illustrated in Algorithm~\ref{algo:st-executor}. It consists of four main phases: (i) exchange of halo regions amongst neighboring processes through non-blocking communications; (ii) execution of core tiles (in overlap with communication); (iii) wait for the termination of the communications; (iv) execution of boundary tiles. 

As explained in Sections~\ref{sec:examples:execution}, a sufficiently deep halo region enables correct computation of the boundary tiles. Further, tiles are executed atomically, meaning that all iterations in a tile are computed without ever synchronizing with other processes. The depth of the boundary region, which affects the amount of off-process data to be redundantly computed, increases with the number $n$ of loops to be fused. In the example in Figure~\ref{fig:st-mpi-init}, there are $n=3$ loops, and three ``strips'' of extra vertices are necessary for correctly computing the fused loops without tile-to-tile synchronizations.

We recall from Section~\ref{sec:tiling:lc} that the {\em depth} of the loop chain indicates the extent of the boundary region. This parameter imposes a limit to the number of fusible loops. If $\mathbb{L}$ includes more loops than the available boundary region -- that is, if $n > \text{{\em depth}}$ --  then $\mathbb{L}$ will have to be split into shorter loop chains, to be fused individually. As we shall see (Section~\ref{sec:implementation}), in our inspector/executor implementation the {\em depth} is controlled by the Firedrake's DMPlex module.


\section{Implementation: SLOPE, PyOP2, and Firedrake}
\label{sec:implementation}

The implementation of automated sparse tiling is distributed over three open-source software modules. 
\begin{description}
\item[Firedrake] An established framework for the automated solution of PDEs through the finite element method~\citep{firedrake}.
\item[PyOP2] A module used by Firedrake to apply numerical kernels over sets of mesh components. Parallelism is handled at this level.
\item[SLOPE] A library for writing inspector/executor schemes, with primary focus on sparse tiling. PyOP2 uses SLOPE to apply sparse tiling to loop chains. 
\end{description}

\begin{figure}[htpb]
\centering
\includegraphics[scale=0.6]{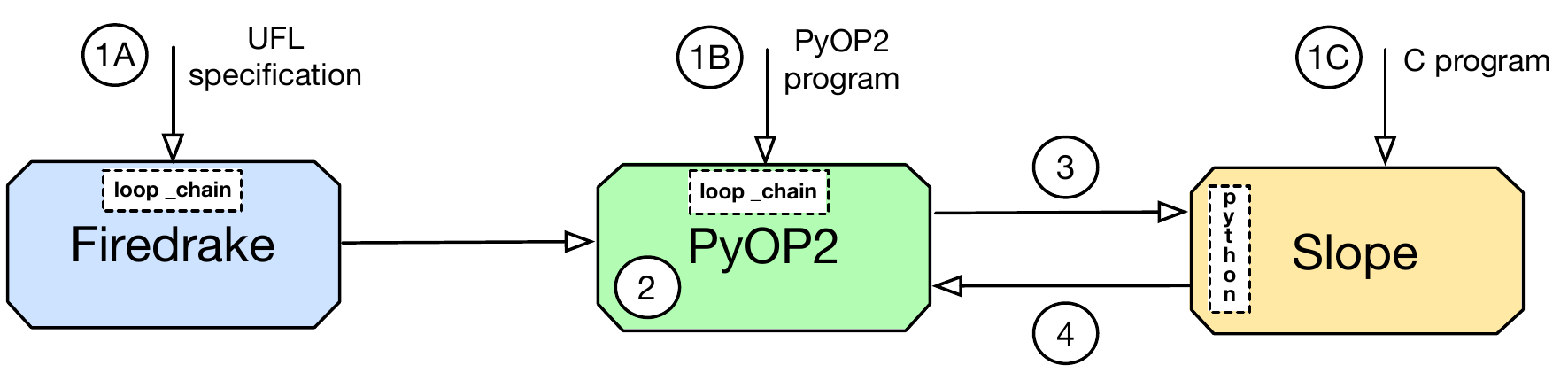}
\caption{Sparse tiling in the Firedrake-PyOP2-SLOPE framework. There are three ways of sparse tiling a loop chain: decorating a Firedrake program (1A), decorating a sequence of loops in a PyOP2 program (1B), writing both the loop chain and the inspector/executor codes explicitly in C through calls to SLOPE (1C). Both (1A) and (1B) use the {\em loop$\_$chain} interface (details in Section~\ref{sec:tiling:lcinterface}). PyOP2 derives the loop chain (2), essentially sets, maps and loops (see Section~\ref{sec:tiling:lc} and example in Figure~\ref{code:tiling-inspector}), from the kernels produced within the {\em loop$\_$chain} context. The loop chain is provided to SLOPE through its Python interface (3). SLOPE performs the inspection and returns its output, a tiles schedule, to PyOP2 (4). Eventually, the executor is generated and run by PyOP2.}
\label{fig:st-implementation}
\end{figure}

The SLOPE library is an open source embodiment of the algorithms presented in this article. The interplay amongst Firedrake, PyOP2 and SLOPE is outlined in Figure~\ref{fig:st-implementation} and discussed in more detail in the following sections.

\subsection{SLOPE: a Library for Sparse Tiling Irregular Computations}
\label{sec:tiling:impl-slope}
SLOPE is an open source software that provides an interface to build loop chains and to express inspector/executor schemes for sparse tiling\footnote{SLOPE is available at \url{https://github.com/coneoproject/SLOPE}}. 

The loop chain abstraction implemented by SLOPE has been formalized in Section~\ref{sec:tiling:lc}. In essence, a loop chain comprises some sets (including the separation into core, boundary, and non-exec regions), maps between sets, and a sequence of loops. Each loop has one or more descriptors specifying what and how different sets are accessed. The example in Figure~\ref{code:tiling-inspector} illustrates the interface exposed by the library. SLOPE implements Algorithms~\ref{algo:st-inspector},~\ref{algo:st-projection} and~\ref{algo:st-tiling} from Section~\ref{sec:algorithm}. Further, it provides additional features to estimate the effectiveness and to verify the correctness of sparse tiling, including generation of VTK files (suitable for Paraview~\citep{paraview}), to visualize the partitioning of the mesh into colored tiles, as well as insightful inspection summaries (showing, for example, number and average size of tiles, total number of colors used, time spent in critical inspection phases).

In the case of shared-memory parallelism, the following sections of code are parallelized through OpenMP:
\begin{description}
\item[Inspection] The projection and tiling algorithms; in particular, the loops over set elements in Algorithm~\ref{algo:st-projection} and Algorithm~\ref{algo:st-tiling}).
\item[Execution] The computation of same colored tiles; that is, the two loops over tiles in Algorithm~\ref{algo:st-executor}.
\end{description}

\subsection{PyOP2: Lazy Evaluation and Interfaces}
\label{sec:tiling:lcinterface}
PyOP2 is a Python library offering abstractions to model an unstructured mesh -- in terms of {\em sets} (e.g. vertices, edges), {\em maps} between sets (e.g., a map from edges to vertices to express the mesh topology), and {\em datasets} associating data to each set element (e.g. 3D coordinates to each vertex) -- and applying numerical kernels to sets of entities~\citep{firedrake}. In this section, we focus on the three relevant contributions to PyOP2 made through our work: (i) the interface to identify loop chains; (ii) the lazy evaluation mechanism that allows loop chains to be built; (iii) the interaction with SLOPE to automatically build and execute inspector/executor schemes.

To apply sparse tiling to a sequence of loops, the {\em loop$\_$chain} interface was added to PyOP2. This interface, exemplified in Figure~\ref{code:loop-chain-interface}, is also exposed to the higher layers, for example in Firedrake. In the listing, the {\tt name} uniquely identifies a loop chain. Other parameters (most of them optional) are useful for performance evaluation and performance tuning. Amongst them, the most important are the {\tt tile$\_$size} and the {\tt fusion$\_$scheme}. The {\tt tile$\_$size} specifies the initial average size for the seed partitions. The {\tt fusion$\_$scheme} allows to specify how to break a long sequence of loops into smaller loop chains, which makes it possible to experiment with a full set of sparse tiling strategies without having to modify the source code.

\begin{figure}[htpb]
\begin{CenteredBox}
\begin{lstlisting}[basicstyle=\scriptsize\ttfamily,morekeywords={with}]
with loop_chain(name, tile_size, fusion_scheme, ...):
    <some PyOP2 parallel loops are expressed/generated here>
\end{lstlisting}
\end{CenteredBox}
\caption{The {\em loop$\_$chain} interface in PyOP2.}
\label{code:loop-chain-interface}
\end{figure}

PyOP2 exploits lazy evaluation of parallel loops to generate an inspector/executor scheme. The parallel loops encountered during the program execution -- or, analogously, those generated through Firedrake -- are pushed into a queue, instead of being executed immediately. The sequence of parallel loops in the queue is called the {\em trace}. If a dataset $f$ needs to be read, for example because a user wants to inspect its values or a global linear algebra operation is performed, then the trace is traversed -- from the most recent parallel loop to the oldest one -- and a new sub-trace produced. The sub-trace includes all parallel loops that must be executed to evaluate $f$ correctly. The sub-trace can then be executed or further pre-processed.

All loops in a trace that were created within a {\em loop$\_$chain} scope are sparse tiling candidates. In detail, the interaction between PyOP2 and SLOPE is as follows:
\begin{enumerate}
\item Figure~\ref{code:loop-chain-interface} shows that a {\em loop$\_$chain} defines a new scope. As this scope is entered, a stamp $s_1$ of the trace is generated. This happens ``behind the scenes'', because the {\em loop$\_$chain} is a Python context manager, which can execute pre-specified routines prior and after the execution of the body. As the {\em loop$\_$chain}'s scope is exited, a new stamp $s_2$ of the trace is computed. All parallel loops in the trace generated between $s_1$ and $s_2$ are placed into a sub-trace for pre-processing.
\item The pre-processing consists of two steps: (i) ``simple'' fusion -- consecutive parallel loops iterating over the same iteration space that do not present indirect data dependencies are merged; (ii) generation of a loop chain representation for SLOPE.
\item In (ii), PyOP2 inspects the sequence of parallel loops and translates their metadata (sets, maps, loops) into a format suitable for the SLOPE's Python interface. SLOPE performs an inspection for the received loop chain and returns a tiles schedule to PyOP2 (i.e., it runs Algorithm~\ref{algo:st-inspector}).
\item A ``software cache'' mapping {\em loop$\_$chain}s to {\em inspection}s is used. This whole process needs therefore be executed only once for a given {\em loop$\_$chain}. 
\item The executor is generated, compiled and run directly by PyOP2, with the help of an API provided by SLOPE. To run the executor, the tiles schedule produced in (3) is used.
\end{enumerate}

\subsection{Firedrake/DMPlex: the S-depth Mechanism for Extended Halo Regions}
\label{sec:tiling:impl-firedrake}
Firedrake uses DMPlex~\citep{dmplex-cite} to handle meshes. DMPlex is responsible for partitioning, distributing over multiple processes, and locally reordering a mesh. The MPI parallelization is therefore managed through Firedrake/DMPlex.

During the start-up phase, each MPI process receives a contiguous partition of the original mesh from DMPlex. The required PyOP2 sets, which can represent either topological components (e.g., cells, vertices) or function spaces, are created. As intuitively shown in Figure~\ref{fig:sets}, these sets distinguish between multiple regions: core, owned, exec, and non-exec. Firedrake initializes the four regions exploiting the information provided by DMPlex. 

To support the loop chain abstraction, Firedrake must be able to allocate arbitrarily deep halo regions. Both Firedrake and DMPlex have been extended to support this feature~\citep{Knepley2015}. A parameter called {\em S-depth} (the name has historical origins, see for instance~\cite{s-depth-paper}) regulates the extent of the halo regions. A value {\em S-depth} $=n$ indicates the presence of $n$ strips of off-process data elements in each set. The default value is {\em S-depth} $=1$, which enables computation-communication overlap when executing a single loop at the price of a small amount of redundant computation along partition boundaries. This is the default execution model in Firedrake.

\section{Performance Evaluation}
\label{sec:performance}


\subsection{The Seigen Computational Framework}
Seigen is a seismological modelling framework capable of solving the elastic wave equation on unstructured meshes. Exploiting the well-known velocity-stress formulation~\citep{Seigen-3}, the seismic model is expressible as two first-order linear PDEs, which we refer to as {\tt velocity} and {\tt stress}. These governing equations are discretized in space through the discontinuous-Galerkin finite element method. The evolution over time is obtained by using a fourth-order explicit leapfrog scheme based on a truncated Taylor series expansion of the velocity and stress fields. The particular choice of spatial and temporal discretizations has been shown to be non-dissipative~\citep{Seigen-1}. More details can be found in~\cite{Seigen-paper}. Seigen, which is built on top of Firedrake, is part of OPESCI, an ecosystem of software for seismic imaging based on automated code generation~\citep{opesci-project}. 

Seigen has a set of test cases, which differ in various aspects, such as the initial conditions of the system and the propagation of waves. However, they are all based upon the same seismological model; from a computational viewpoint, this means that, in a time step, the same sequence of loops is executed. In the following, we focus on the {\tt explosive$\_$source} test case (see the work by \cite{Garvin1956} for background details).

\subsection{Implementation and Validation}
In a time loop iteration, eight linear systems need to be solved, four from {\tt velocity} and four from {\tt stress}. Each solve consists of three macro-steps: assembling a global matrix $A$; assembling a global vector $b$; computing $x$ in the system $Ax = b$. There are two global ``mass'' matrices, one for {\tt velocity} and one for {\tt stress}. Both matrices are time invariant, so they are assembled before entering the time loop, and block-diagonal, as a consequence of the spatial discretization employed (a block belongs to an element in the mesh). The inverse of a block-diagonal matrix is again block-diagonal and is determined by computing the inverse of each block. The solution of the linear system $Ax = b$, expressible as $x = b A^{-1}$, can therefore be evaluated by looping over the mesh and computing a ``small'' matrix-vector product in each element, where the matrix is a block in $A^{-1}$. Assembling the global vectors boils down to executing a set of loops over mesh entities, particularly over cells, interior facets, and exterior facets. Overall, twenty-five loops are executed in a time loop iteration. Thanks to the hierarchy of ``software caches'' employed by Firedrake, the translation from mathematical syntax into loops is only performed once. 

Introducing sparse tiling into Seigen was relatively straightforward. Three steps were required: (i) embedding the time stepping loop in a {\em loop$\_$chain} context (see Section~\ref{sec:tiling:lcinterface}), (ii) propagating user input relevant for performance tuning, (iii) creating a set of {\em fusion schemes}. A fusion scheme establishes which sub-sequences of loops within a {\em loop$\_$chain} will be fused and the respective seed tile sizes. If no fusion schemes were specified, all of the twenty-five loops would be fused using a default tile size. As we shall see, operating with a set of small loop chains and heterogeneous tile sizes is often more effective than fusing long sequences of loops. 

\begin{figure}[t]
\centering
\begin{subfigure}[b]{0.33\textwidth}
\includegraphics[width=\textwidth]{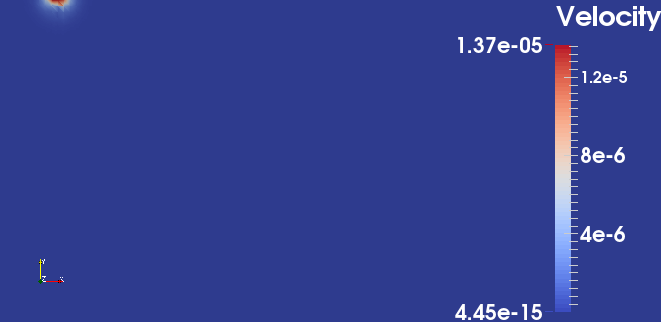}
\caption{Right after the explosion.}
\end{subfigure}%
~~~
\begin{subfigure}[b]{0.33\textwidth}
\includegraphics[width=\textwidth]{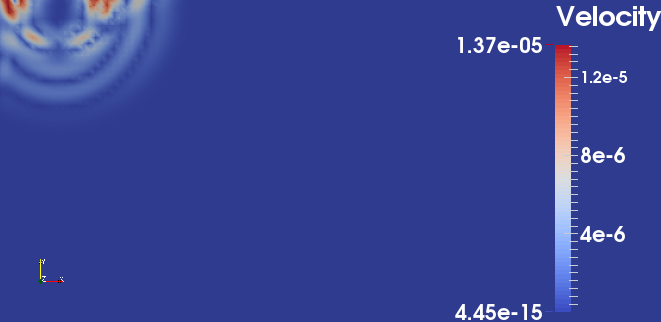}
\caption{Wave propagation.}
\end{subfigure}%
~~~
\begin{subfigure}[b]{0.33\textwidth}
\includegraphics[width=\textwidth]{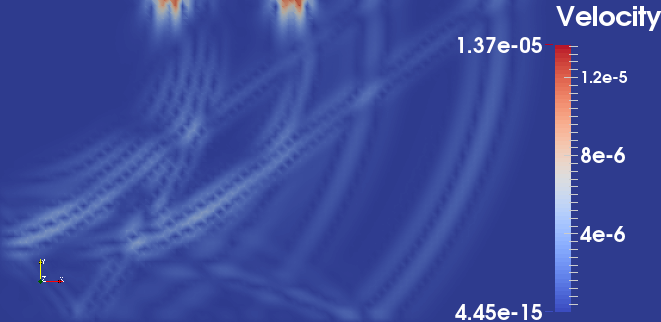}
\caption{Final snapshot.}
\end{subfigure}%

\caption{Three phases of the {\tt explosive$\_$source} test case in Seigen, following an explosion at a point source.}
\label{fig:seigen-output}
\end{figure}

Seigen has several mechanisms to validate the correctness of the seismological model and the test cases. The numerical results of all code versions (with and without tiling) were checked and compared. Paraview was also used to verify the simulation output. Snapshots of the simulation output are displayed in Figure~\ref{fig:seigen-output}.

\subsection{Computational Analysis of the Loops}
We here discuss computational aspects of the twenty-five fusible loops. The following considerations derive from an analytical study of the data movement in the loop chain, extensive profiling through the Intel VTune Amplifier tool~\citep{vtune}, and roofline models (available in~\cite{Seigen-paper}).

Our initial hypothesis was that Seigen would have benefited from sparse tiling. Not only does data reuse arise within single loops (e.g., by accessing vertex coordinates from adjacent cells), but also across consecutive loops, through indirect data dependencies. This seemed to make Seigen a natural fit for sparse tiling. The eight ``solver'' loops perform matrix-vector multiplications in each mesh element. It is well established that linear algebra operations of this kind are memory-bound. Four of these loops arise from {\tt velocity}, the others from {\tt stress}. There is significant data reuse amongst the four {\tt velocity} loops and amongst the four {\tt stress} loops, since the same blocks in the global inverse matrices are accessed. We therefore hypothesized performance gains if these loops were fused through sparse tiling.

We also observed that because of the particular mesh employed, the exterior facet loops, which implement the boundary conditions of the variational problems, have negligible cost. However, the cells and facets loops do have significant cost, and data reuse across them arises. Six loops iterate over the interior mesh facets to evaluate facet integrals, which ensure the propagation of information between adjacent cells in discontinuous-Galerkin methods. The operational intensity of these loops is much lower than that of cell loops, and memory-boundedness is generally expected. Consecutive facet and cell integral loops share fields, which creates cross-loop data reuse opportunities, thus strengthening the hypothesis about the potential of sparse tiling in Seigen. 

All computational kernels generated in Seigen are optimized through COFFEE~\citep{Luporini-coffee}, a system used in Firedrake that, in essence, (i) minimizes the operation count by restructuring expressions and loop nests, and (ii) maximizes auto-vectorization opportunities by applying transformations such as array padding and by enforcing data alignment.

\subsection{Setup and Reproducibility}
\label{sec:performance:setup}
There are two parameters that we can vary in {\tt explosive$\_$source}: the polynomial order of the method, $q$, and the input mesh. We test the spectrum $q \in \lbrace 1, 2, 3, 4 \rbrace$. To test higher polynomial orders, changes to both the spatial and temporal discretizations would be necessary. For the spatial discretization, the most obvious choice would be tensor product function spaces, which at the moment of writing is still unavailable in Firedrake. We use as input a two-dimensional rectangular domain of fixed size 300$\times$150 tessellated with unstructured triangles ({\tt explosive$\_$source} only supports two-dimension domains); to increase the number of elements in the domain, thus allowing to weak scale, we vary the mesh spacing $h$.

The generality of the sparse tiling algorithms, the flexibility of the {\em loop$\_$chain} interface, and the {\em S-depth} mechanism made it possible to experiment with a variety of fusion schemes without changes to the source code.  Five fusion schemes were devised, based on the following criteria: (i) amount of data reuse, (ii) amount of redundant computation over the boundary region, (iii) memory footprint (the larger, the smaller the tile size to fit in cache). The fusion schemes are summarized in Table~\ref{table:seigen-fusion-schemes}. The full specification, along with the seed tile size for each sub loop chain, is available at~\cite{fabio_luporini_2017_840000}.

The experimentation was conducted on two platforms, whose specification is reported in Table~\ref{table:seigen-setup}. The two platforms, Erebus (the ``development machine'') and Helen (a Broadwell-based architecture in the Helen cluster~\citep{cx2-helen}) were idle and exclusive access had been obtained when the runs were performed. Support for shared-memory parallelism is discontinued in Firedrake, so only distributed-memory parallelism with 1 MPI process per physical core was tested. The MPI processes were pinned to cores. The hyperthreading technology was tried, but found to be generally non-profitable. The code used for running the experiments was archived with the Zenodo data repository service: Firedrake~\citep{lawrence_mitchell_2017_836680}, PETSc~\citep{barry_smith_2017_836685}, PETSc4py~\citep{lisandro_dalcin_2017_836684}, FIAT~\citep{miklos_homolya_2017_836679}, UFL~\citep{martin_sandve_alnaes_2017_836683}, TSFC~\citep{miklos_homolya_2017_836677}, PyOP2~\citep{florian_rathgeber_2017_836688}, COFFEE~\citep{fabio_luporini_2017_836678}, SLOPE~\citep{fabio_luporini_2017_836738}, and Seigen~\citep{fabio_luporini_2017_840000}.

\begin{table}[htpb]
\tiny
\centering
\begin{minipage}[t]{.60\textwidth}
{
\begin{tabulary}{1.2\textwidth}{M{1.75cm} | M{2.65cm} | M{3.0cm} N}
\hline
System & Erebus & Helen \\
\hlineB{4}
Node & \shortstack{1x4-core\\Intel I7-2600 3.4GHz} & \shortstack{2x14-core\\Intel Xeon E5-2680 v4 2.40GHz} & \\[6pt] \hline
DRAM & 16 GB & 128 GB (node) &\\[6pt] \hline
Cache hierarchy & L1=32KB, L2=256KB, L3=8MB & L1=32KB, L2=256KB, L3/socket=35MB & \\[6pt] \hline
Compilers & Intel {\tt icc} 16.0.2 & Intel {\tt icc} 16.0.3 &\\[6pt] \hline
Compiler flags & {\tt -O3 -xHost -ip} & {\tt -O3 -xHost -ip} & \\[6pt] \hline
MPI version & Open MPI 1.6.5 & SGI MPT 2.14 &\\ \hline
\end{tabulary}
}
\caption{Systems specification.}
\label{table:seigen-setup}
\end{minipage}\hfill
\begin{minipage}[t]{.38\textwidth}
\makebox[\textwidth][c]
{
\begin{tabulary}{1.0\columnwidth}{M{0.5cm} | M{1.1cm} | M{1.6cm} | M{0.65cm} N}
\hline
Fusion scheme & Number of loop chains & Criterion & {\em S-depth}  \\
\hlineB{4}
{\tt fs1} & 3 & Fuse costly cells and facets loops & 2 & \\[5pt] \hline
{\tt fs2} & 8 & More aggressive than {\tt fs1} & 2 & \\[5pt] \hline
{\tt fs3} & 6 & {\tt fs2}, include all solver loops & 2 & \\[5pt] \hline
{\tt fs4} & 3 & More aggressive than {\tt fs3} & 3 & \\[5pt] \hline
{\tt fs5} & 2 & {\tt velocity}, {\tt stress} & 4 & \\[5pt] \hline
\end{tabulary}
}
\caption{Fusion schemes summary.}
\label{table:seigen-fusion-schemes}
\end{minipage}\hfill

\end{table}

In the following, for each experiment, we collect three measurements.

\begin{description}
\item[Overall completion time -- OT] Used to compute the maximum application speed-up when sparse tiling is applied. 
\item[Average compute time -- ACT] Sparse tiling impacts the kernel execution time by increasing data locality. Communication is also influenced, especially in aggressive fusion schemes: the rounds of communication decrease, while the data volume exchanged may increase. ACT isolates the gain due to increased data locality from (i) the communication cost and (ii) any other action performed in Firedrake (executing Python code) between the invocation of kernels. This value is averaged across the processes.
\item[Average compute and communication time -- ACCT] As opposed to ACT, the communication cost is also included in ACCT. By comparing ACCT and ACT, the communication overhead can be derived. 
\end{description}
As we shall see, all of these metrics will be essential for a complete understanding of the sparse tiling performance. 

To collect OT, ACT and ACCT, the following configuration was adopted. All experiments were executed with ``warm cache''; that is, with all kernels retrieved directly from the Firedrake's software cache of compiled kernels, so code generation and compilation times are not counted. All of the non-tiled {\tt explosive$\_$source} tests were repeated three times. The minimum times are reported (negligible variance). The cost of global matrix assembly -- an operation that takes place before entering the time loop -- {\it is not} included in OT. Firedrake needs to be extended to assemble block-diagonal matrices directly into vectors (an entry in the vector would represent a matrix block). Currently, this is instead obtained in two steps: first, by assembling into a CSR matrix; then, by explicitly copying the diagonal into a vector (a Python operation). The assembly per se never takes more than 3 seconds, so it was reasonable to exclude this temporary overhead from our timing. The inspection cost due to sparse tiling {\it is} included in OT, and its overhead will be discussed appropriately in a later section. Extra costs were minimized: no check-pointing, only two I/O sessions (at the beginning and at the end of the computation), and minimal logging. The time loop has a fixed duration, while the time step depends on the mesh spacing $h$ to satisfy the Courant-Friedrichs-Lewy necessary condition (i.e., CFL limit) for the numerical convergence. In essence, finer meshes require proportionately smaller time steps to ensure convergence.

We now proceed with discussing the achieved results. Below, a generic instance of {\tt explosive$\_$source} optimized with sparse tiling will be referred to as a ``tiled version'', otherwise the term ``original version'' will be used.

\subsection{Single-node Experimentation}
\label{sec:performance:singlenode}
Hundreds of single-node experiments were executed on Erebus, which was easily accessible in exclusive mode and much quicker to use for VTune profiling. The rationale of these experiments was to assess how sparse tiling impacts the application performance by improving data locality.

We only show parallel runs at maximum capacity (1 MPI process per physical core); the benefits of sparse tiling in sequential runs tend to be negligible (if any), because (i) the memory access latency is only marginally affected when a large proportion of bandwidth is available to a single process, (ii) hardware prefetching is impaired by the small iteration space of tiles, (iii) translation lookaside buffer (TLB) misses are more frequent due to tile expansion. Points (ii) and (iii) will be elaborated upon.

Table~\ref{table:seigen-speedups} reports execution times and speed-ups, indicated with the symbol $\Omega$, of the tiled version over the original version for the three metrics OT, ACT and ACCT. We report the best speed-up obtained after varying a number of parameters (tile size, fusion scheme and other optimizations discussed below).

\begin{table}
\tiny
\makebox[\textwidth][c]
{
\begin{tabulary}{1.0\columnwidth}{c V{4} c V{4} M{2.0cm} | M{2.0cm} V{4} M{1.2cm} | M{1.2cm} | M{1.2cm} N}
\hline
$h$ & $q$ & OT original (s) & OT tiled (s) & $\Omega^{OT}$ & $\Omega^{ACT}$ & $\Omega^{ACCT}$ \\
\hlineB{4}
\multirow{4}{*}{$1.0$} & 1 & 687 & 596 & 1.15 &  1.17 & 1.16 \\ 
& 2 & 1453 & 1200 & 1.21 & 1.25 & 1.25 \\ 
& 3 & 3570 & 2847 & 1.25 & 1.28 & 1.27 \\
& 4 & 7057 & 5827 & 1.21 & 1.22 & 1.22 \\
\hline
\multirow{4}{*}{$1.2$} & 1 & 419 & 365 & 1.15 & 1.17 & 1.16 \\ 
& 2 & 870 & 715 & 1.22 & 1.26 & 1.26 \\ 
& 3 & 1937 & 1549 & 1.25 & 1.28 & 1.27\\
& 4 & 4110 & 3390 & 1.21 & 1.23 & 1.22 \\
\end{tabulary}
}
\caption{OT, ACT and ACCT on Erebus, with 1 MPI process per core.}
\label{table:seigen-speedups}
\end{table}

The parameters that we empirically varied to obtain these results were: (i) the fusion scheme, {\tt fs$X$, $X \in \lbrace 1, 2, 3, 4, 5\rbrace$}, see Table~\ref{table:seigen-fusion-schemes}; (ii) the seed tile size -- for each {\tt fs} and $q$, up to four different values chosen to maximize the likelihood of fitting in L2 or L3 cache, were tried\footnote{A less extensive experimentation with ``more adverse'' tile sizes showed that: (i) a very small value causes dramatic slow-downs (up to 8$\times$ slower than the original versions); (ii) larger values cause proportionately greater performance drops.}. 

Further, a smaller experimentation varying (i) type of indirection maps (local or global, see Section~\ref{sec:algorithm}) and (ii) tile shape (through different mesh partitioning algorithms), as well as introducing (iii) software prefetching and (iv) extended boundary region (to minimize redundant computation) led to the conclusion that these optimizations may either improve or worsen the execution time, in ways that are too difficult to predict beforehand. We therefore decided to exclude these parameters from the full search space. This simplifies the analysis that will follow and also allowed the execution of the whole test suite in less than five days. The interested reader is invited to refer to the Supplementary Materials attached to this article for more information.

\begin{figure}[t]
\centering
\begin{subfigure}[b]{0.25\textwidth}
\includegraphics[width=\textwidth]{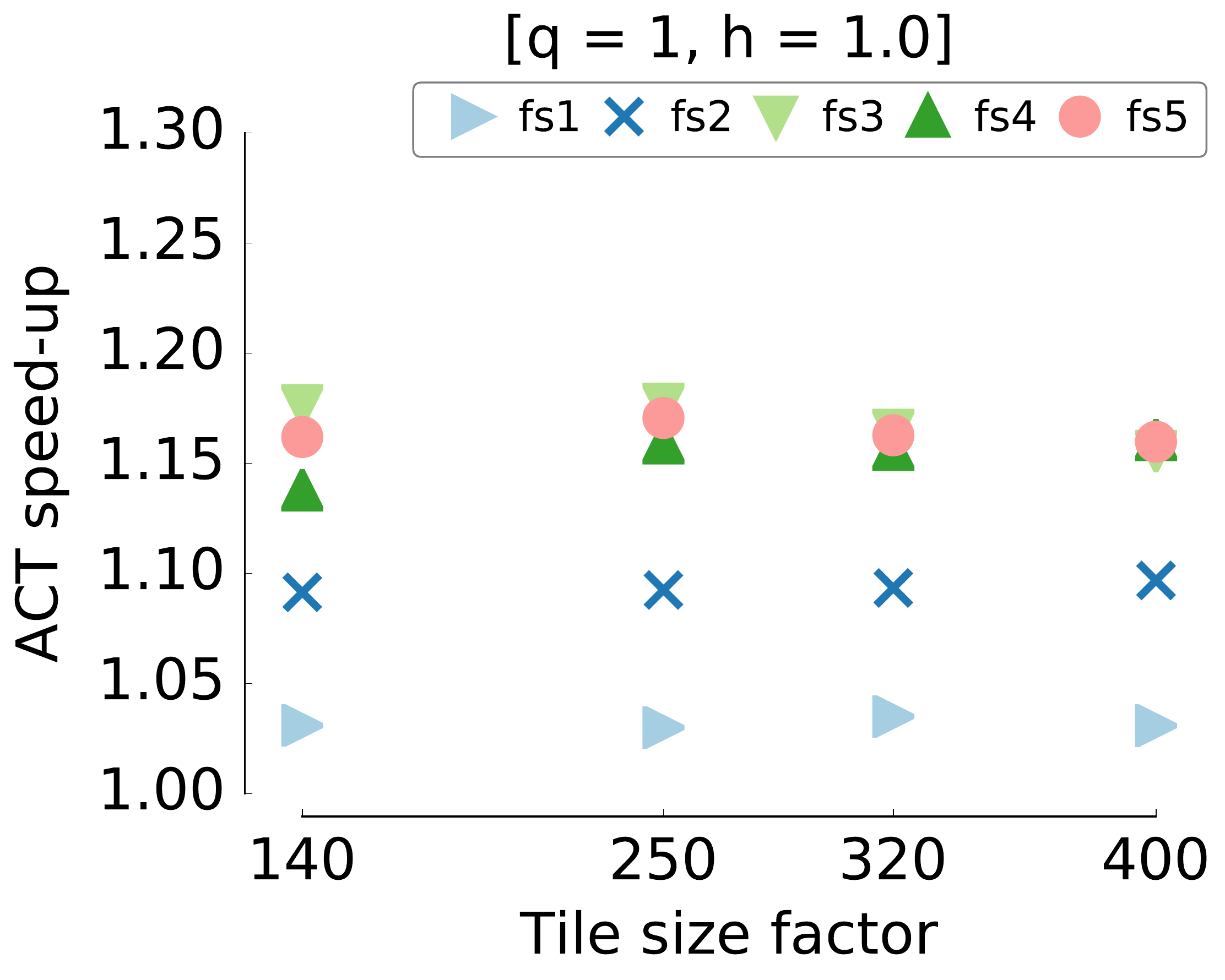}
\end{subfigure}%
~~
\begin{subfigure}[b]{0.25\textwidth}
\includegraphics[width=\textwidth]{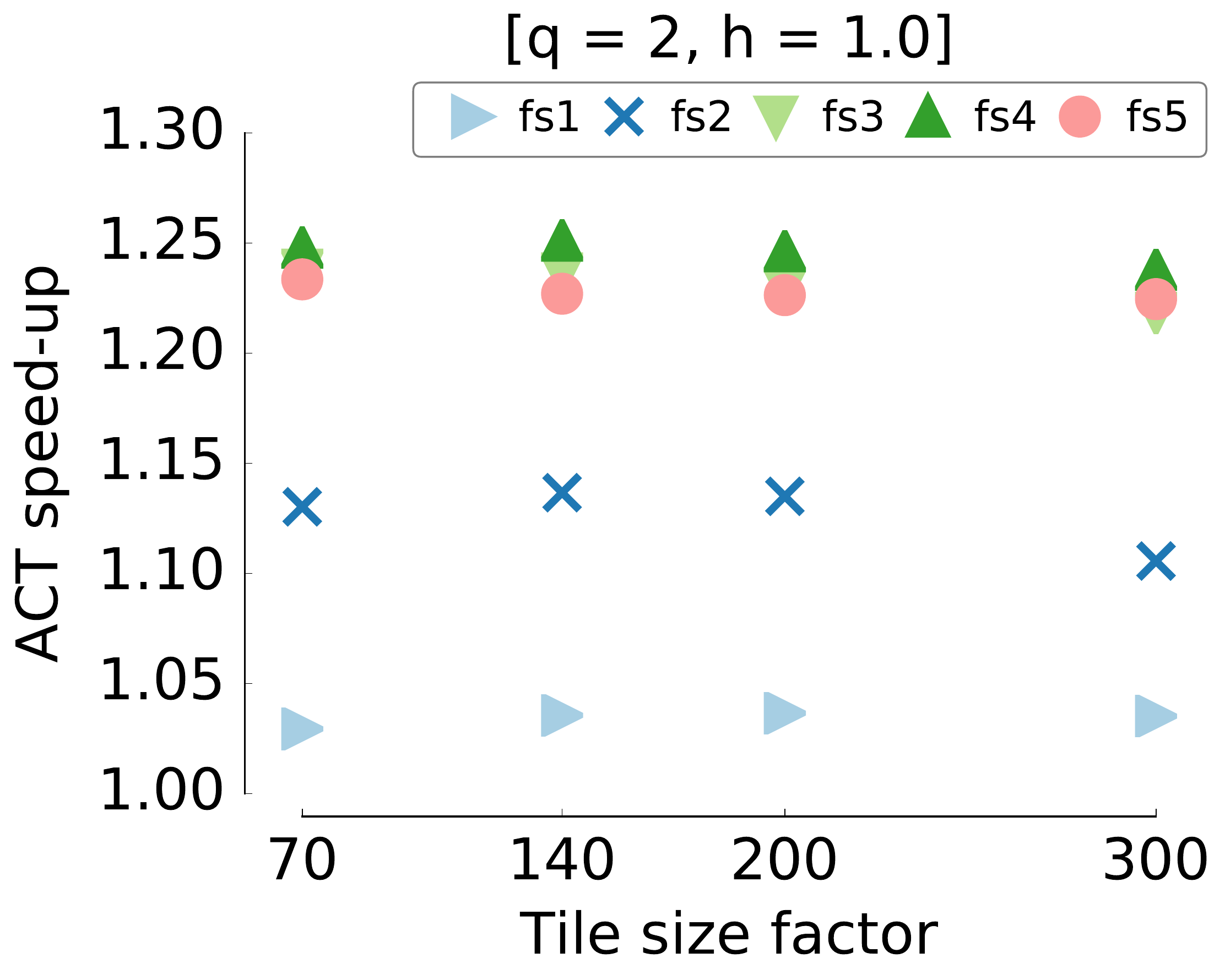}
\end{subfigure}%
~~
\begin{subfigure}[b]{0.25\textwidth}
\includegraphics[width=\textwidth]{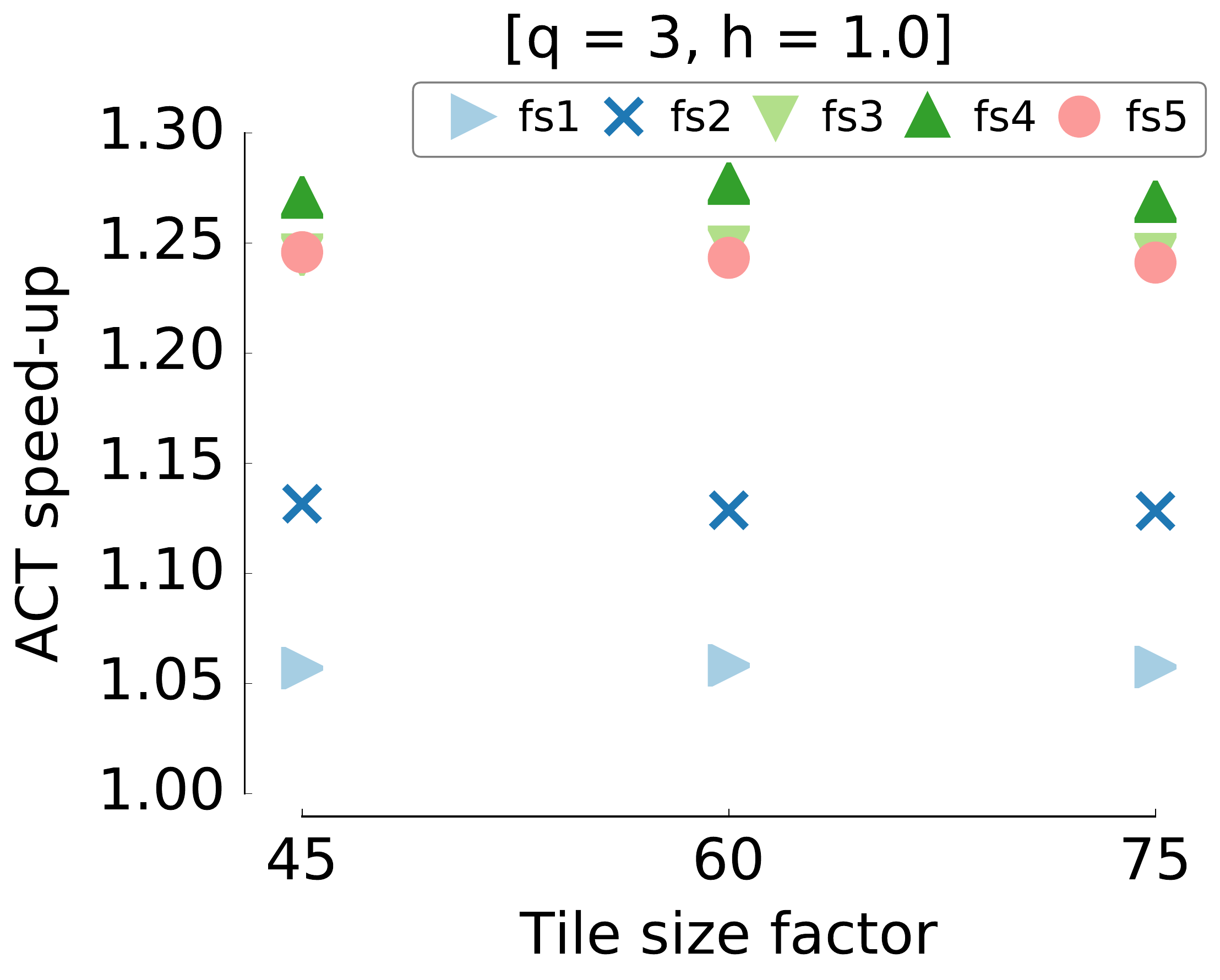}
\end{subfigure}%
~~
\begin{subfigure}[b]{0.25\textwidth}
\includegraphics[width=\textwidth]{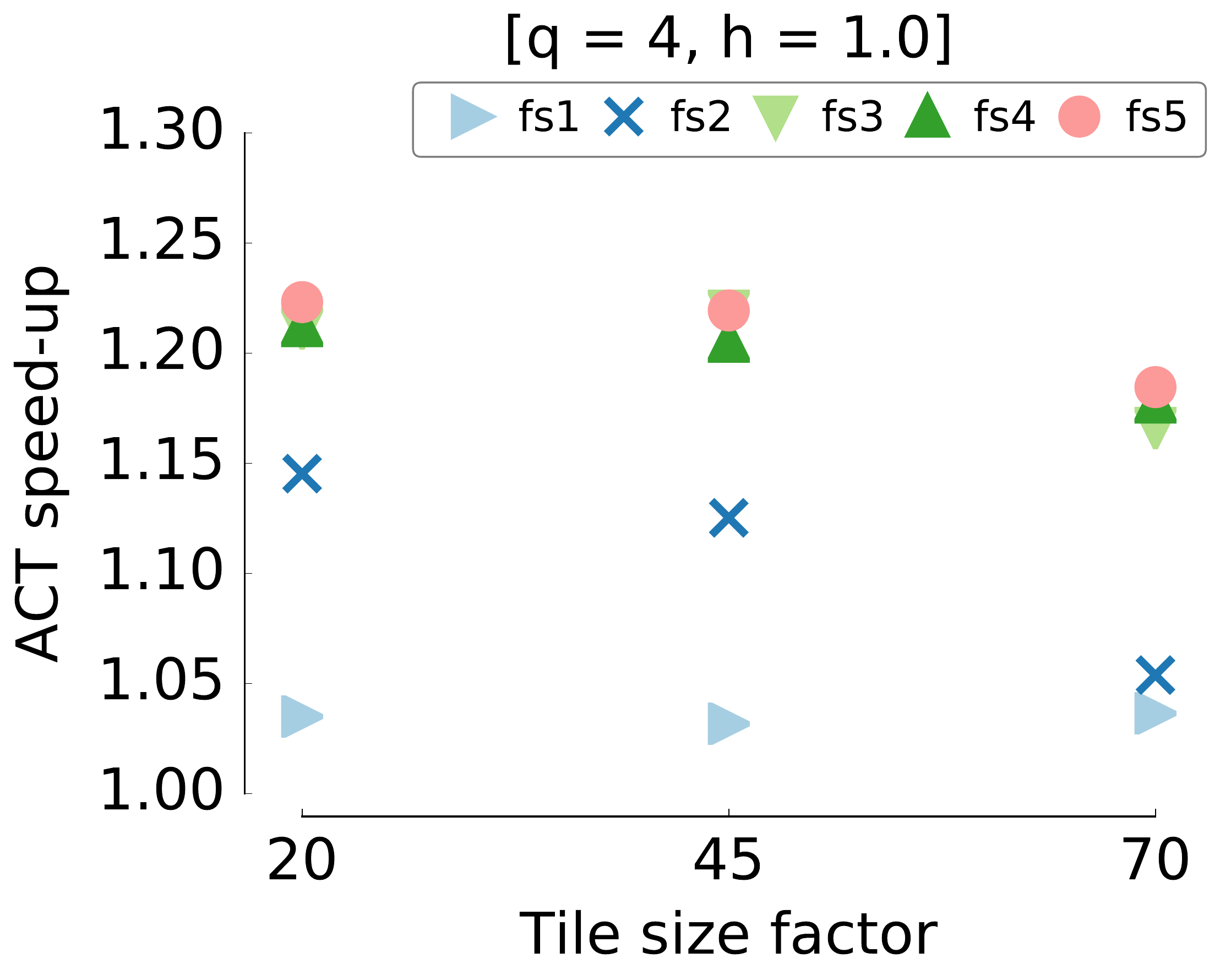}
\end{subfigure}%
~\\
~\\
\begin{subfigure}[b]{0.25\textwidth}
\includegraphics[width=\textwidth]{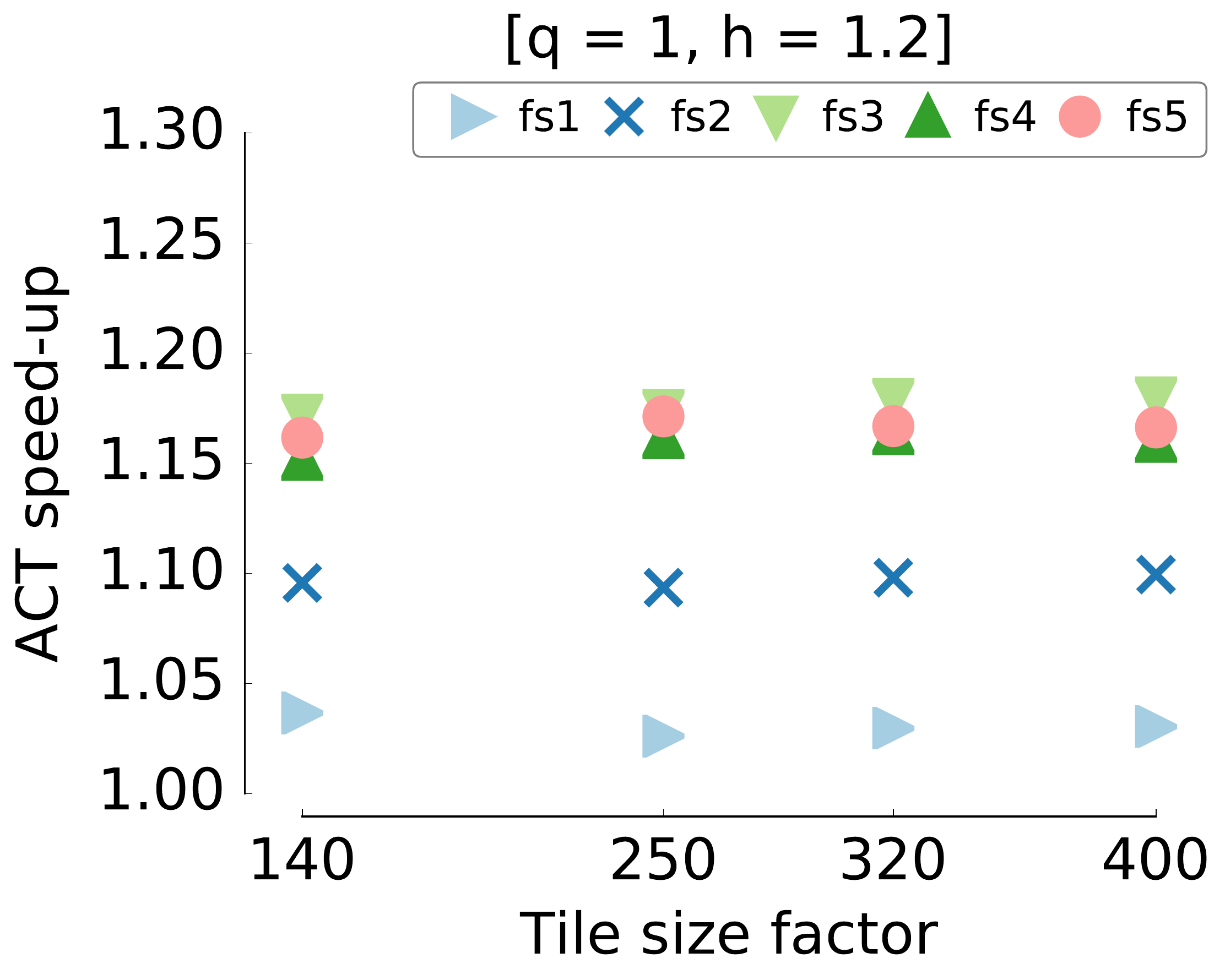}
\end{subfigure}%
~~
\begin{subfigure}[b]{0.25\textwidth}
\includegraphics[width=\textwidth]{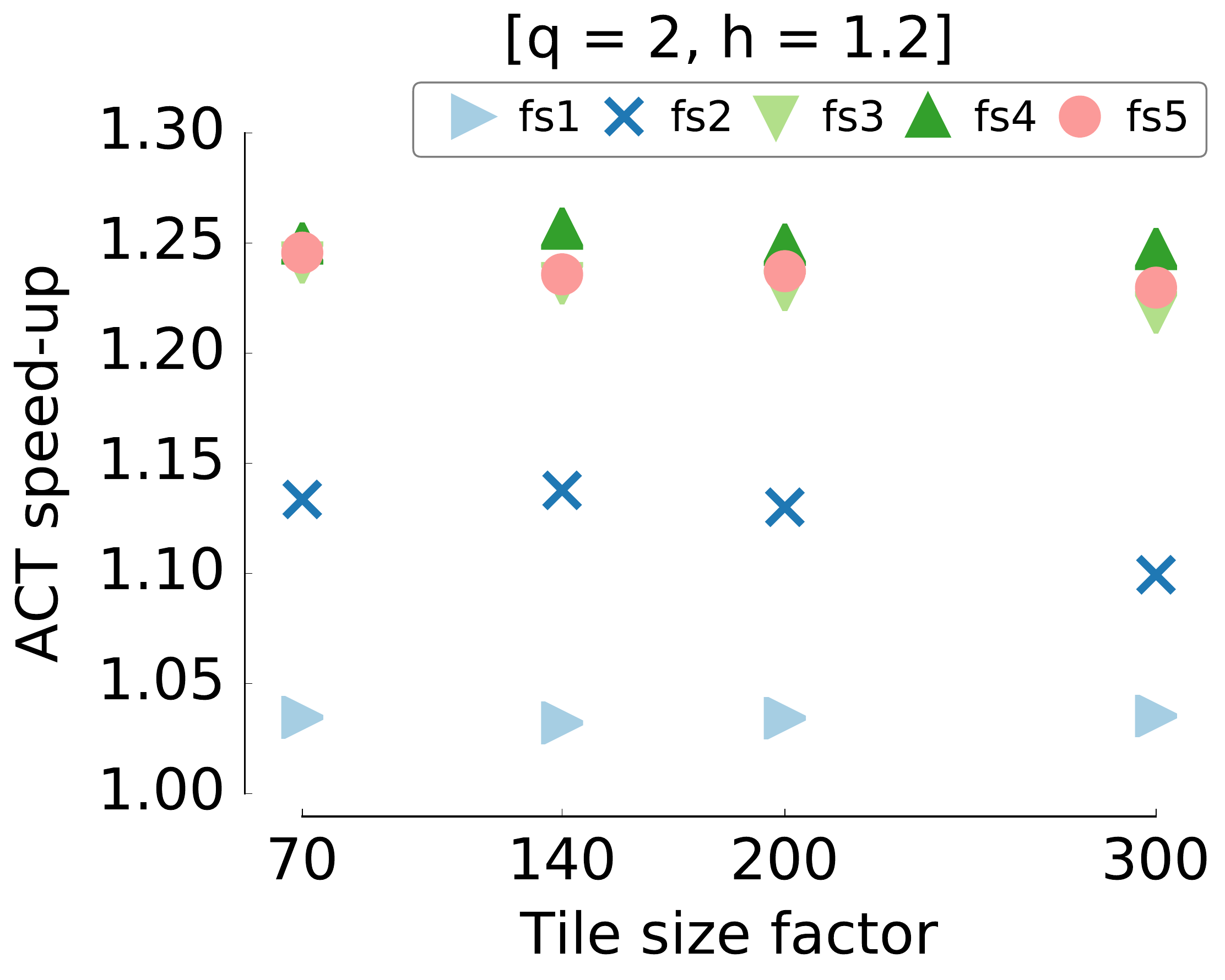}
\end{subfigure}%
~~
\begin{subfigure}[b]{0.25\textwidth}
\includegraphics[width=\textwidth]{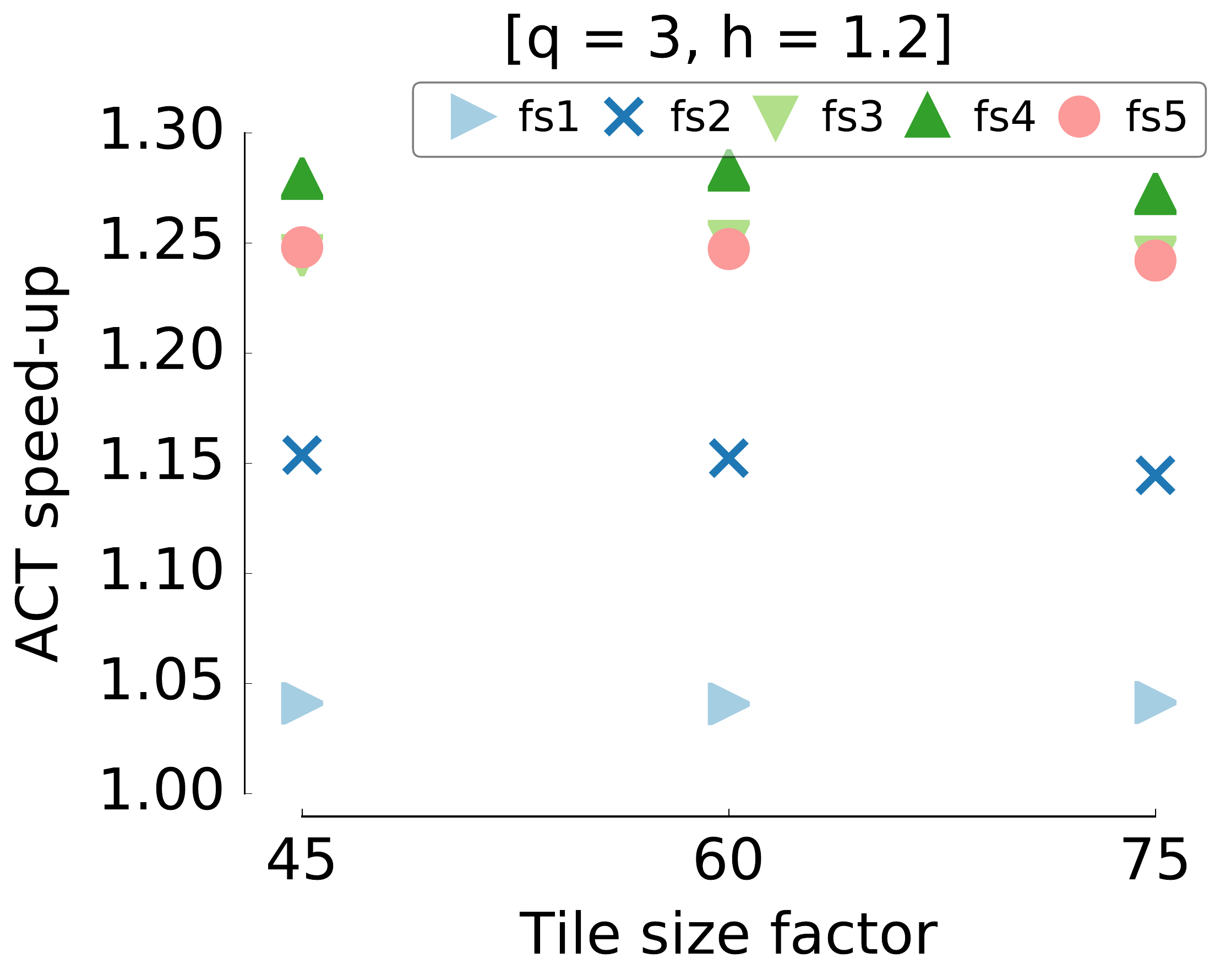}
\end{subfigure}%
~~
\begin{subfigure}[b]{0.25\textwidth}
\includegraphics[width=\textwidth]{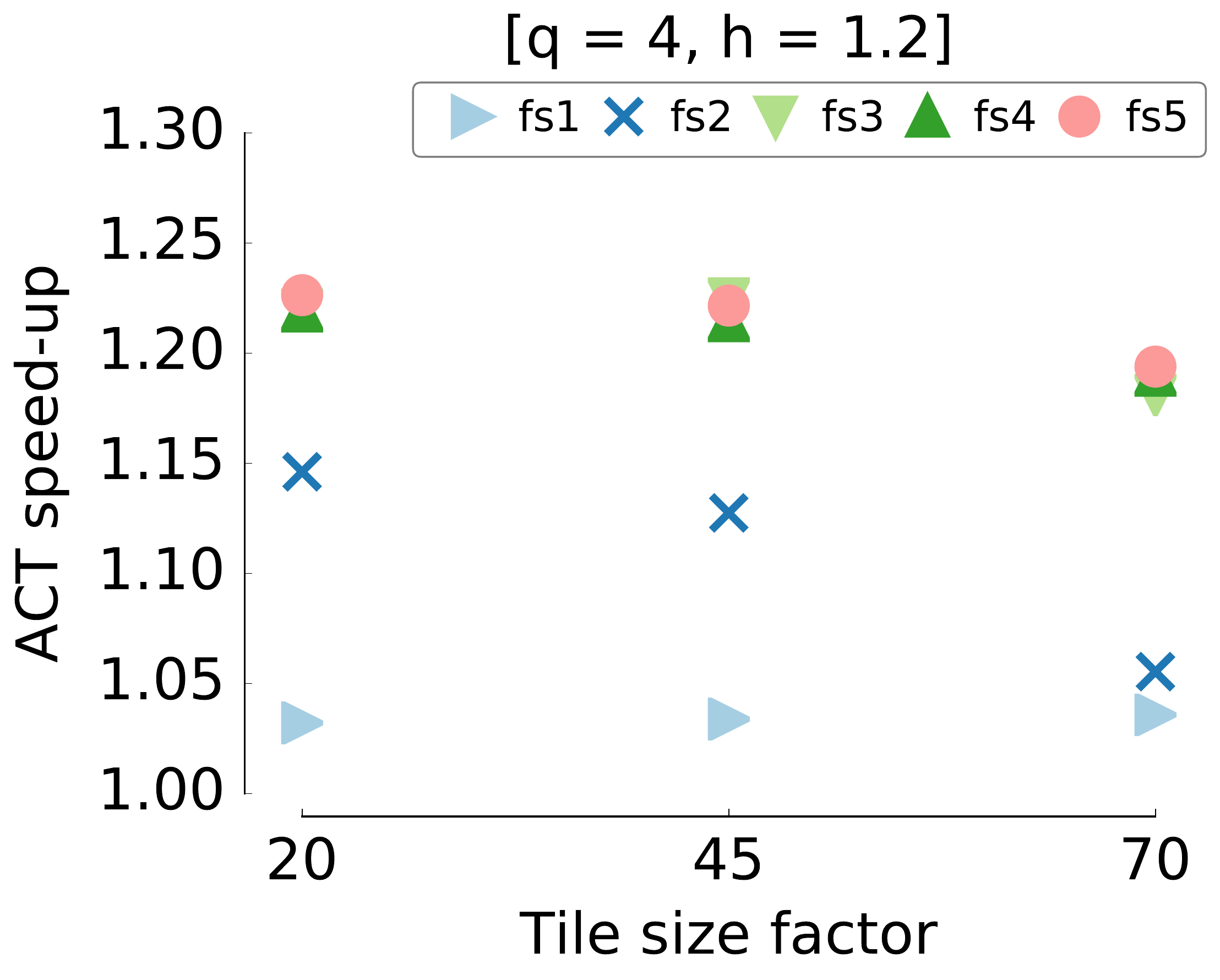}
\end{subfigure}%
\caption{Search space exploration on Erebus, with $h \in \lbrace 1.0, 1.2 \rbrace$ and $q \in \lbrace 1, 2, 3 ,4\rbrace$. Each plot shows the average compute time (ACT) speed-up achieved by multiple tiled versions over the original (non-tiled) version. The seed tile size of a loop chain in an {\tt fs}, in terms of seed loop iterations, is the product of the ``tile size factor'' (x-axis) and a pre-established multiplier (an integer in the range $[1, 4]$; full list available at~\cite{fabio_luporini_2017_840000}).}
\label{fig:st-erebus-expl}
\end{figure}

In Figure~\ref{fig:st-erebus-expl}, we summarize the results of the search space exploration. A plot shows the ACT speed-ups achieved by multiple tiled versions over the non-tiled version for given $q$ and $h$. In these single-node experiments, the ACCT trend was basically identical (as one can infer from Table~\ref{table:seigen-speedups}), with variations in speed-up smaller than 1$\%$.

PyOP2 was enhanced with a \textit{loop chain analyzer} (LCA) capable of estimating the best- and worst-case tile memory footprint, as well as the percentage of data reuse ideally achievable\footnote{It is part of our future plans to integrate this tool with SLOPE, where the effects of tile expansion can be taken into account to provide better estimates.}. We use this tool, as well as Intel VTune, to explain the results shown in Figure~\ref{fig:st-erebus-expl}. We make the following observations.

\begin{itemize}
\item {\tt fs}, unsurprisingly, is the parameter having largest impact on the ACT. By influencing the fraction of data reuse convertible into data locality, the amount of redundant computation and the data volume exchanged, fusion schemes play a fundamental role in sparse tiling. This makes automation much more than a desired feature: without any changes to the source code, multiple sparse tiling strategies could be studied and tuned. Automation is one of our major contributions, and this performance exploration justifies the implementation effort. 
\item There is a non-trivial relationship between ACT, $q$ and {\tt fs}. The aggressive fusion schemes are more effective with high $q$ -- that is, with larger memory footprints -- while they tend to be less efficient, or even deleterious, when $q$ is low. The extreme case is {\tt fs5}, which fuses two long sequences of loops (twelve and thirteen loops each). In Figure~\ref{fig:st-erebus-expl} (Erebus), {\tt fs5} is never a winning choice, although the difference between {\tt fs3}/{\tt fs4} and {\tt fs5} decreases as $q$ grows. If this trend continued with $q > 4$, then the gain from sparse tiling could become increasingly larger.
\item A non-aggressive scheme fuses only a few small subsets of loops. As discussed in later sections, these fusion schemes, despite affording larger tile sizes than the more aggressive ones (due to the smaller memory footprint), suffer from limited cross-loop data reuse. For {\tt fs1}, LCA determines that the percentage of reusable data in the three fused loop chains decreases from 25$\%$ ($q=1$) to 13$\%$ ($q=4$). The drop is exacerbated by the fact that no reuse can be exploited for the maps. Not only are these ideal values, but also a significant number of loops are left outside of loop chains. The combination of these factors motivate the lack of substantial speed-ups. With {\tt fs2}, a larger proportion of loops are fused and the amount of shared data increases. The peak ideal reuse in a loop chain reaches 54$\%$, which translates into better ACTs. A similar growth in data reuse can be appreciated in more aggressive fusion schemes, with a peak of 61$\%$ in one of the {\tt fs5}'s loop chains.  Nevertheless, the performance of {\tt fs5} is usually worse than {\tt fs4}. As we shall clarify in Section~\ref{sec:performance:limit}, this is mainly due to the excessive memory footprint, which in turn leads to very small tiles. Speculatively, we tried running a sixth fusion scheme: a single loop chain including all of the 25 loops in a time iteration. In spite of an ideal data reuse of about 70$\%$, the ACT was always significantly higher than all other schemes.
\item Figure~\ref{fig:st-erebus-expl} shows the ACT for a ``good selection'' of tile size candidates. Our approach was as follows. We took a very small set of problem instances  and we tried a large range of seed tile sizes. Very small tile sizes caused dramatic slow-downs, mostly because of ineffective hardware prefetching and TLB misses. Tile sizes larger than a certain fraction of L3 cache (usually slightly larger than what a core should ideally own) also led to increasingly higher ACTs. If we consider $q=4$ in Figure~\ref{fig:st-erebus-expl}, we observe that the ACT of {\tt fs2}, {\tt fs3}, and {\tt fs4} grows when the initial number of iterations in a tile is as big as 70. Here, LCA shows that the tile memory footprint is, in multiple loop chains, higher than 3MB, with a peak of 6MB in {\tt fs3}. This exceeds the proportion of L3 cache that a process owns (on average), which explains the performance drop. 
\end{itemize}

\subsection{Multi-node Experimentation}
\label{sec:performance:multinode}
Weak-scaling experiments were carried out on thirty-two nodes split over two racks in the Helen cluster at Imperial College London~\citep{cx2-helen,helen-top}. The node architecture as well as the employed software stack are summarized in Table~\ref{table:seigen-setup}. The rationale of these experiments was to assess whether the change in communication pattern and amount of redundant computation caused by sparse tiling could affect the run-time performance.

\begin{figure}[htpb]
\centering
\includegraphics[scale=0.3]{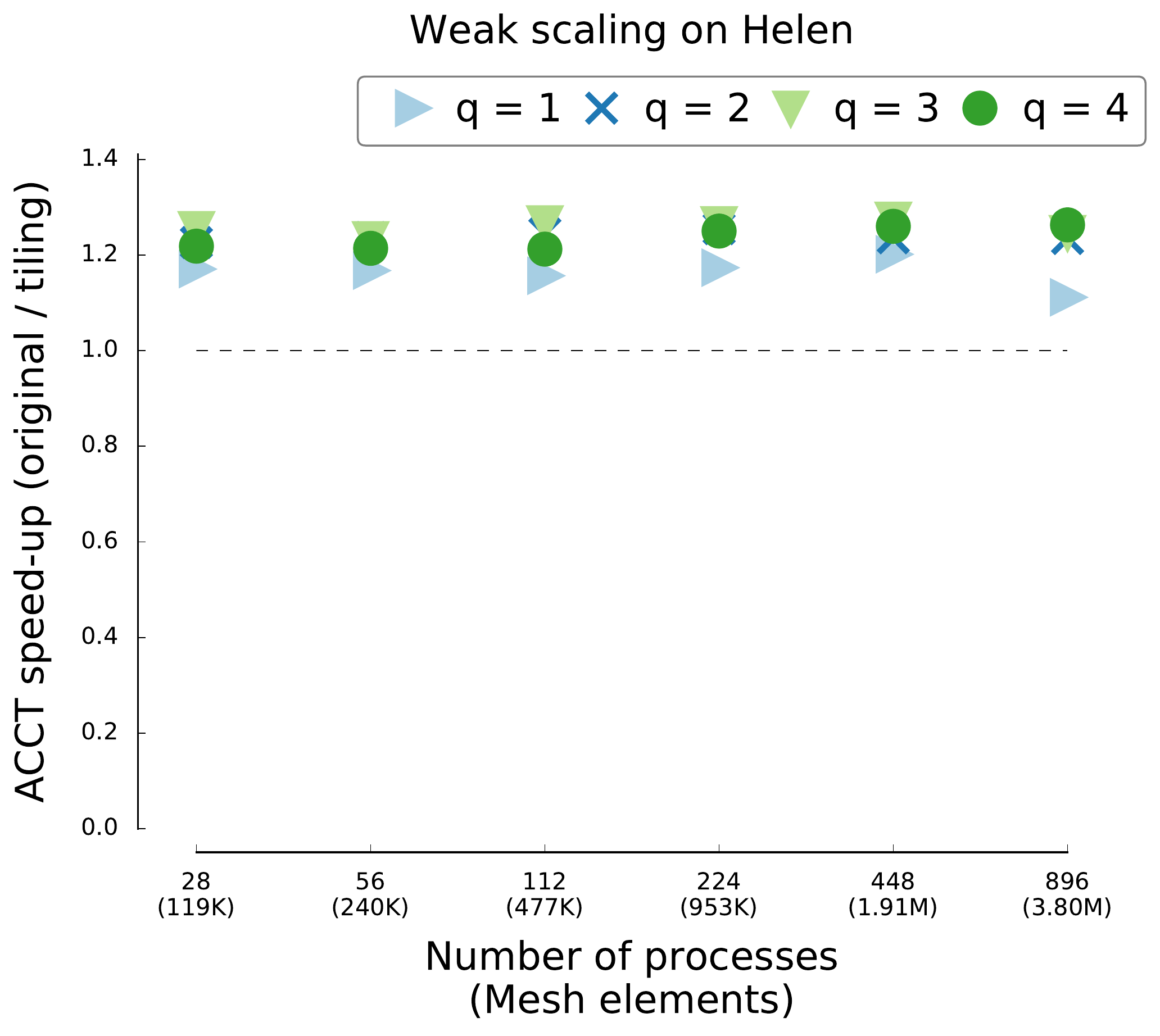}
\captionof{figure}{Weak scaling performance of Seigen's {\tt explosive$\_$source} on the Helen cluster. The ACCT speed-up is relative to a single process. Results for $q \in \lbrace 1, 2, 3 ,4\rbrace$ are displayed. The simulations ran for 1000 timesteps.}
\label{fig:st-helen}
\end{figure}

For each $q$ in the usual range $[1-4]$, {\tt fs3} generally resulted in the best performance improvements, due to its trade-off between gained data locality and restrained redundant computation (whose effect obviously worsen as $q$ grows). Figure~\ref{fig:st-helen} summarizes the ACCT speed-ups achieved by the best tiled version (i.e., the one found by empirically varying the tile size, with same tile size factor as in Figure~\ref{fig:st-erebus-expl}) over the original version. The weak scaling trend is remarkable, with only a small drop in the case $q=1$ when switching from one rack (448 cores) to two racks (896 cores), which disappears as soon as the per-process workload becomes significant. For instance, with $q = 2$, each process already computes over more than 150k degrees of freedom for the velocity and stress fields. The peak speed-up over the original version, 1.28$\times$, was obtained in the test case $q = 3$ when running with 448 processes. The performance achieved was 1.84 TFLOPs/s (158 GFLOPs required at each time step, for a total of 1000 time steps and an ACCT of 86 seconds); this corresponds to roughly 15$\%$ of the theoretical machine peak (assuming AVX base frequency; the architecture ideally performs 16 double-precision FLOPs per cycle).

\subsection{Negligible Inspection Overhead}
As explained in Section~\ref{sec:performance:setup}, the inspection cost was included in OT. In this section, we quantify this overhead in a representative problem instance. In all other problem instances, the overhead was either significantly smaller than or essentially identical (for reasons discussed below) to that reported here. We consider {\tt explosive$\_$source} on Erebus with $h=1.0$, $q=4$, and {\tt fs5}. With this configuration, the time step was $\Delta t=481 \cdot 10^{-6}$ (we recall from Section~\ref{sec:performance:setup} that $\Delta t$ is a function of $h$). Given the simulation duration $T=2.5$, in this test case 5198 time steps were performed. A time step took on average 1.15 seconds. In each time step, twenty-five loops, fused as specified by {\tt fs5}, are executed. We know that in {\tt fs5} there are two sub loop chains, which respectively consist of thirteen and twelve loops. To inspect these two loop chains, 1.4 and 1.3 seconds were needed (average across the four MPI ranks, with negligible variance). Roughly 98$\%$ of the inspection time was due to the projection and tiling functions, while only 0.2$\%$ was spent on the tile initialization phase (see Section~\ref{sec:algorithm}). These proportions are consistent across other fusion schemes and test cases. After 200 time steps (less than 4$\%$ of the total) the inspection overhead was already close to 1$\%$. Consequently, the inspection cost, in this test case, was eventually negligible. The inspection cost increases with the number of fused loops, which motivates the choice of {\tt fs5} for this analysis. 

\subsection{On the Main Performance Limiting Factor}
\label{sec:performance:limit}
The tile structure, namely its shape and size, is the key factor affecting sparse tiling in Seigen. 

\begin{figure}[t]
\centering
\begin{subfigure}[b]{0.33\textwidth}
\includegraphics[width=\textwidth]{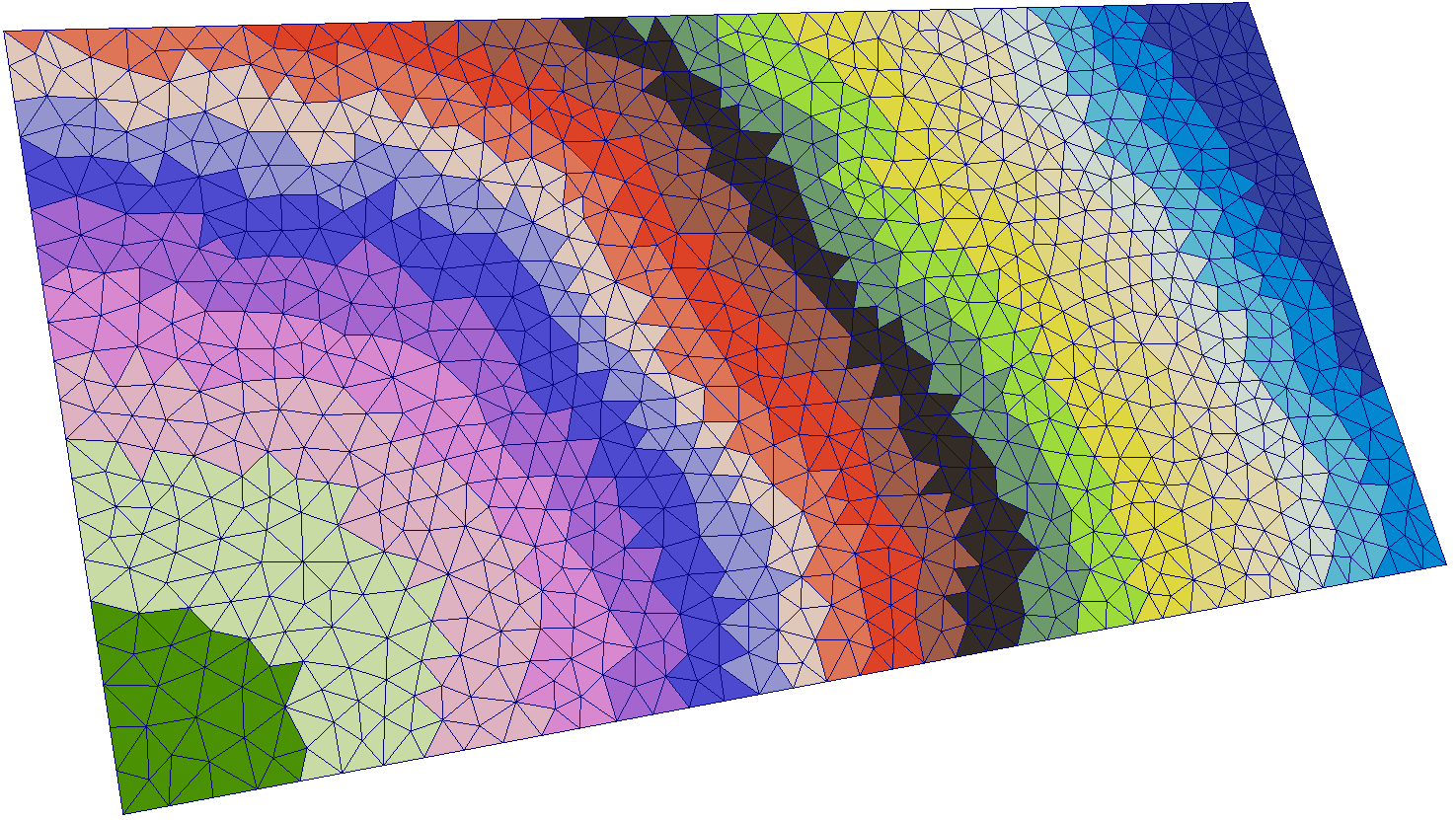}
\caption{{\tt chunk} partitioning in Seigen.}
\label{fig:tile-shapes:chunk}
\end{subfigure}%
~~
\begin{subfigure}[b]{0.33\textwidth}
\includegraphics[width=\textwidth]{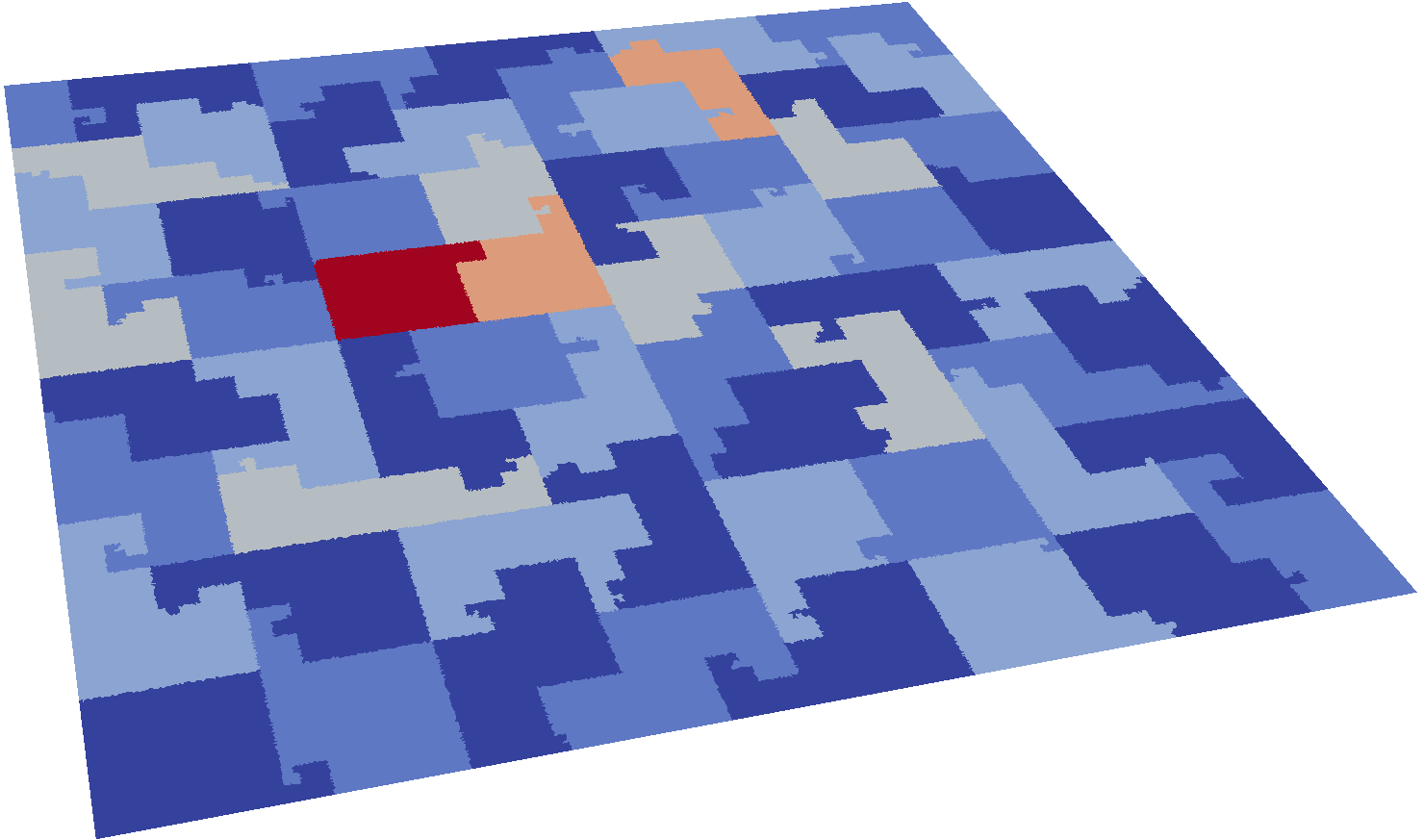}
\caption{Ideal {\tt Hilbert} partitioning.}
\label{fig:tile-shapes:hilbert}
\end{subfigure}%
~~
\begin{subfigure}[b]{0.33\textwidth}
\includegraphics[width=\textwidth]{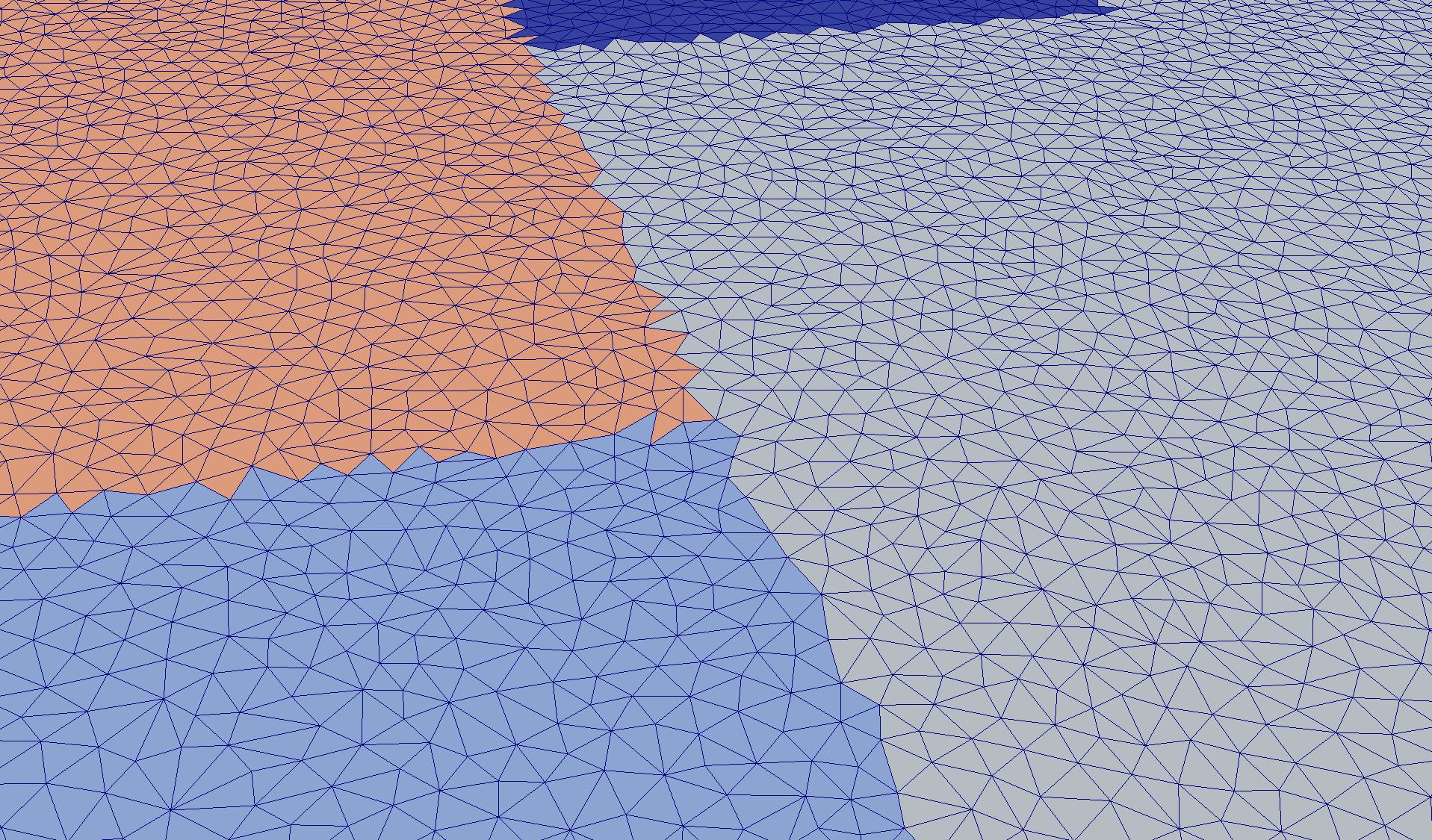}
\caption{Zoomed in {\tt Hilbert} partitioning.}
\label{fig:tile-shapes:hilbert-zoomed}
\end{subfigure}%
\label{fig:tile-shapes}
\caption{Representation of seed iteration space partitioning strategies via SLOPE and Paraview.}
\end{figure}

The seed loop partitioning and the mesh numbering determine the tile structure. The simplest way of creating tiles consists of ``chunking'' the seed iteration space every $\mathrm{ts}$ iterations, with $\mathrm{ts}$ being the initial tile size (see Section~\ref{sec:algorithm}). Hence, the {\tt chunk} partitioning inherits the original mesh numbering. In Firedrake, and therefore in Seigen, meshes are renumbered during initialization applying the reverse Cuthill-McKee (RCM) algorithm. Using {\tt chunk} partitioning on top of an RCM-renumbered mesh has the effect of producing thin, rectangular tiles, as displayed in Figure~\ref{fig:tile-shapes:chunk}. This dramatically affects tile expansion, as a large proportion of elements will lie on the tile border. There are potential solutions to this problem. The most promising would consist of using a Hilbert curve, rather than RCM, to renumber the mesh. This would lead to more regular polygonal tiles when applying {\tt chunk} partitioning. Figures~\ref{fig:tile-shapes:hilbert} and~\ref{fig:tile-shapes:hilbert-zoomed} show the tile structure that we would ideally want in Seigen, from a Hilbert-renumbered mesh produced outside of Firedrake. As later discussed, introducing support for Hilbert curves in Firedrake is part of our future work.

The memory footprint of a tile grows quite rapidly with the number of fused loops. In particular, the matrices accessed in the {\tt velocity} and {\tt stress} solver loops have considerable size. The larger the memory footprint, the smaller $\mathrm{ts}$ for the tile to fit in some level of cache. Allocating small tiles has unfortunately multiple implications. First, the proportion of iterations lying on the border grows, which worsen the tile expansion phenomenon discussed, thus affecting data locality. Secondly, a small $\mathrm{ts}$ impairs hardware prefetching, since the virtual address streams become more irregular. Finally, using small tiles implies that a proportionately larger number of tiles are needed to ``cover'' the iteration space; in the case of shared-memory parallelism, this increases the probability of coloring conflicts, which result in higher inspection costs.

As reiterated in the previous sections, the maximum length of a loop chain is dictated by the extent of the boundary region. Unfortunately, a long loop chain has two undesirable effects. First, the redundant computation overhead tends to be non-negligible if some of the fused loops are compute-bound. Sometimes, the overhead could be even larger than the gain due to increased data locality. Second, the size of an S-level (a ``strip'' of boundary elements) grows with the depth of the boundary region, as outer levels include more elements than inner levels. This increases not only the amount of redundant computation, but also the volume of data to be communicated. Overall, this suggests that multiple, short loop chains may guarantee the best performance improvement, as indicated by the results in Sections~\ref{sec:performance:singlenode} and~\ref{sec:performance:multinode}.

\section{Related Work}
\label{sec:related-work}

The data dependence analysis that we have developed in this article is based on the \textit{loop chain} abstraction, which was originally presented in~\cite{ST-KriegerHIPS2013}. This abstraction is sufficiently general to capture data dependencies in programs structured as arbitrary sequences of loops, particularly to create inspector/executor schemes for many unstructured mesh application. Inspector/executor strategies were first formalized by~\cite{ST-Saltz91}. They have been used to exploit data reuse and to expose shared-memory parallelism in several studies~\citep{ST-dimeEtna00,ST-StroutLCPC2002,ST-Demmel08,ST-KriegerIAAA2012}.

Sparse tiling is a technique based upon inspection/execution. The term was coined by~\cite{ST-StroutLCPC2002,ST-StroutIJHPCA} in the context of the Gauss-Seidel algorithm and also used in~\cite{ST-StroutPLDI03} in the Moldyn benchmark. However, the technique was initially proposed by~\cite{ST-dimeEtna00} to parallelize computations over unstructured meshes, taking the name of \textit{unstructured cache blocking}. In this work, the mesh was initially partitioned; the partitioning represented the tiling in the first sweep over the mesh. Tiles would then shrink by one layer of vertices for each iteration of the loop. This shrinking represented what parts of the mesh could be accessed in later iterations of the outer loop without communicating with the processes executing other tiles. The unstructured cache blocking technique also needed to execute a serial clean-up tile at the end of the computation.~\cite{ST-Adams99c} also developed an algorithm very similar to sparse tiling, to parallelize Gauss-Seidel computations. The main difference between~\cite{ST-StroutLCPC2002,ST-StroutIJHPCA} and~\cite{ST-dimeEtna00} was that in the former work the tiles fully covered the iteration space, so a sequential clean-up phase at the end could be avoided. All these approaches were either specific to individual benchmarks or not capable of scheduling across heterogeneous loops (e.g., one over cells and another over degrees of freedom). These limitations had been addressed in~\cite{st-paper}.

The automated code generation technique presented in \cite{ST-OhioStateMPICodeGen} examines the data affinity among loops and performs partitioning with the goal of minimizing inter-process communication, while maintaining load balancing. This technique supports unstructured mesh applications (being based on an inspector/executor strategy) and targets distributed memory systems, although it does not exploit the loop chain abstraction and does not introduce any sort of loop reordering transformation.

Automated code generation techniques based on polyhedral compilers have been applied to structured grid benchmarks or proxy applications~\cite{pluto}. However, there has been very little effort in providing evidence that these tools can be effective in real-world applications. Time-loop diamond tiling was applied in~\cite{cohen-timetiling} to a proxy application, but experimentation was limited to shared-memory parallelism. In~\cite{reguly-ops-tiling}, instead, an approach based on raising the level of abstraction, similar to the one presented in this paper, is described and evaluated. The experimentation is conducted using realistic stencil codes ported to the OPS library. The main difference with respect to our work is the focus on structured grids (i.e., different types of numerical methods are targeted).

In structured codes, multiple layers of halo, or ``ghost'' elements, are often used to reduce communication~\citep{Bassetti98}. Overlapped tiling exploits the very same idea: trading communication for redundant computation along the boundary~\citep{Zhou12}. Several works tackle overlapped tiling within single regular loop nests (mostly stencil-based computations), for example~\cite{Meng09,Krishnamoorthy07,Chen02}. Techniques known as ``communication avoiding''~\citep{ST-Demmel08,ST-commAvoidingSparse2009} also fall in this category. To the best of our knowledge, overlapped tiling for unstructured mesh applications has only been studied analytically, by~\cite{gihan-overlapped}. Further, we are not aware of any prior techniques for automation. 

\section{Future Work}
Using a Hilbert curve numbering will lead to dramatically better tile shapes, thus mitigating the performance penalties due to tile expansion, TLB misses and hardware prefetching described in Section~\ref{sec:performance}. This extension is at the top of our future work priorities. 

Shared-memory parallelism was not as carefully tested as distributed-memory parallelism. First of all, we would like to replace the current OpenMP implementation in SLOPE with the MPI Shared Memory (SHM) model introduced in MPI-3. Not only does a unified programming model provide significant benefits in terms of maintainability and complexity, but the performance may also be greater as suggested by recent developments in the PETSc community. Secondly, some extra work would be required for a fair comparison of this new hybrid MPI+MPI programming model with and without sparse tiling. 

The experimentation was carried out on a number of ``conventional'' Intel Xeon architectures; we aim to repeat the experimentation on the Intel's Knights Landing soon. 

Finally, a cost model for automatic derivation of fusion schemes and tile sizes is still missing.

\section{Conclusions}
Sparse tiling aims to turn the data reuse in a sequence of loops into data locality. In this article, three main problems have been addressed: the specialization of sparse tiling to unstructured mesh applications via a revisited loop chain abstraction, automation via DSLs, effective support for shared- and distributed-memory parallelism. The major strength of this work lies in the fact that all algorithmic and technological contributions presented derive from an in-depth study of realistic application needs. The performance issues we found through Seigen would never have been exposed by a set of simplistic benchmarks, as often used in the literature. Further experimentation will be necessary when 3D domains and high-order discretizations will be supported by Seigen. In essence, the performance experimentation shows systematic speed-ups, in the range 1.10$\times$-1.30$\times$. This is hopefully improvable by switching to Hilbert curve numberings and by exploiting shared memory through a suitable paradigm. Finally, our opinion is that sparse tiling is an ``extreme'' optimization: at least for unstructured mesh application, it is unlikely that it will lead to speed-ups in the order of magnitudes. However, through automation via DSLs, and with suitable optimization and tuning, it may still play a key role in improving the performance of real-world computations.

\section{Supplementary materials}
\subsection{Summary of the Optimizations Attempted in Seigen}
\label{app:summary-opts}

A number of potential optimizations were attempted when applying sparse tiling to Seigen. The experimentation with these optimizations was, however, inconclusive. Although performance improvements were observed in various problem instances, in a significant number of other cases either a detrimental effect or no benefits were noticed at all. Below we briefly discuss the impact of four different execution strategies: (i) use of global maps, (ii) variation in tile shape, (iii) software prefetching, (iv) use of an extended boundary region to minimize redundant computation.

\begin{description}
\item[Global and local maps] Algorithm~\ref{algo:st-inspector} computes so called local maps to avoid an extra level of indirection in the executor. Although no data reuse is available for the local maps (each fused loop has its own local maps), there might be benefits from improved hardware prefetching and memory latency. We compared the use of global and local maps (i.e., the former are normally constructed by Firedrake and provided in the loop chain specification, the latter are computed by SLOPE), but no definitive conclusion could be drawn, as both performance improvements and deteriorations within a 5$\%$ range were observed.

\item[Tile shape] In Section~\ref{sec:performance:limit} we have explained that a Hilbert-renumbered mesh might substantially improve the tile shape quality. Hilbert curves, however, are not supported in Firedrake yet. An alternative consists of partitioning the seed iteration space with a library like METIS~\citep{METIS} before applying RCM. We experimented and discovered that this approach too was not exempt from side effects. The main cause was increased translation lookaside buffer (TLB) misses, which occur whenever the CPU cannot retrieve the mapping to the physical page corresponding to a given virtual page. Since the page table has a hierarchical structure, handling a TLB miss usually requires multiple accesses to memory. Hence, TLB misses are much more costly than cache misses. Sparse tiling increases the TLB miss/hit ratio because of the fragmented streams of virtual addresses. This is evident (and more pronounced) when the tile size is small, in which case a TLB miss is quite likely to occur when jumping to executing a new loop. This problem is exacerbated by the {\tt metis} partitioning (in contrast to {\tt chunk}), which leads to irregular tile shapes. Here, tile expansion may eventually incorporate iterations living in completely different virtual pages. VTune experimentation with $q=1$ and $q=2$ versions of {\tt explosive$\_$source} showed that {\tt chunk}- and {\tt metis}-based sparse tiling suffer from an increase in TLB misses of roughly 16$\%$ and 35$\%$, respectively. To mitigate this issue, we explored the possibility of using larger virtual pages through Linux's Transparent Huge Pages mechanism, which was enabled to automatically allocate memory in virtual pages of 2MB (instead of the default 4KB) -- as long as the base array addresses were properly aligned. However, no significant differences were observed, and a deeper investigation is still necessary. 

\item[Software prefetching] In a loop, there is usually more than a single stream of memory accesses amenable to hardware prefetching (e.g., accesses to the indirection maps; direct accesses to data values; indirect accesses to data values if the mesh has a good numbering). Sparse tiling, unfortunately, impairs hardware prefetching for two reasons: (i) the virtual addresses streams are considerably shorter; (ii) tile expansion introduces irregularities in these streams. Software prefetching can be used together with hardware prefetching to minimize memory stalls. PyOP2 and SLOPE have been extended to emit intrinsics instructions to prefetch the iteration $i$'s maps and data values while executing the iteration $i-k$ at distance $k$. No compelling evidence that this further transformation could systematically improve the performance was found.

\item[Extended boundary region] The special non-exec tile $T_{ne}$ (see Sections~\ref{sec:examples} and~\ref{sec:algorithm}) reduces the amount of redundant computation in long loop chains by expanding over boundary tiles. There are two ways of creating $T_{ne}$: either an extra layer of data is added to the boundary region (e.g., see Figure~\ref{fig:st-mpi-init}) or during inspection, by searching for mesh boundaries. The current implementation only supports the first option. Manually deriving $T_{ne}$ would be not only algorithmically complex, but also potentially very expensive. 
\end{description}

\begin{printonly}
See the Supplementary Materials in the online version.
\end{printonly}

\bibliographystyle{ACM-Reference-Format}
\bibliography{bibliography}
